%Paper: hep-th/9305020
%From: SPALLUCCI@TRIESTE.INFN.IT
%Date: Thu, 6 May 1993 16:54:33 +0200 (WET-DST)

%%%%%%%%%%%%%%%%%%%%%%%%%%%%%%%%%%%%%%%%%%%%%%%%%%
%        TEX DIALECT: PHYZZX
%%%%%%%%%%%%%%%%%%%%%%%%%%%%%%%%%%%%%%%%%%%%%%%%%%
\def\gb{{\bar g}}
\def\X{{\dot X}}
\def\dis{\displaystyle}
\def\m{membrane}
\def\ms{membranes}
\def\r{\rho}
\def\L{{\cal L}}
\def\P{\Pi_{\lambda\mu\nu}}
\def\G{{\cal G}}
\def\p{\Pi^{\lambda\mu\nu}}

\def\Pcl{\Pi_{{\rm cl.}}^{\mu\nu\rho}}
\def\pcl{\Pi_{{\rm cl.}\,\mu\nu\rho}}
\def\ncl{N_{\rm cl.}}
\def\f{F_{\lambda\mu\nu}}
\def\F{F^{\lambda\mu\nu}}
\def\ff{F_{\mu\nu}}

\def\Y{\dot Y}
\input phyzzx
\hfuzz 30pt
\pubnum{UTS-DFT-92-5}
\titlepage
\singlespace
\title{Gauge Theory of Relativistic Membranes}
\author{Antonio Aurilia\foot{E-Mail address: AAURILIA@CSUPOMONA.EDU}}
\smallskip
\address{Department of Physics\break
California State Polytechnic University\break Pomona, CA 91768}
\smallskip
\andauthor{Euro Spallucci\foot{E-Mail address:
SPALLUCCI@TRIESTE.INFN.IT}}
\smallskip
\address{ Dipartimento di Fisica Teorica \break Universit\`a di
Trieste, \break
INFN, Sezione di Trieste \break Trieste, Italy 34014 }
\vfill
\centerline{Short title: The Membrane Geodesic Field}
\medskip
\centerline{P.A.C.S. : 11.17}
\vfill
\eject

\abstract

A relativistic membrane is usually represented by the Dirac-Nambu-Goto
action in terms of the extremal area of a 3-dimensional timelike
submanifold
of Minkowski space. In this paper we show that a relativistic membrane
admits an equivalent representation in terms of the Kalb-Ramond gauge
field $F_{\mu\nu\rho}=\partial_{\,[\,\mu}B_{\nu\rho]}$ encountered in
string theory. At first glance this is somewhat surprising, since the
Kalb-Ramond field is usually interpreted as the spin-0 radiation field
generated by a closed string.

By `` equivalence '' of the two representations we mean the following:
\hfill\break
if $x=X(\xi)$ is a solution
of the classical equations of motion derived from the Dirac-Nambu-Goto
action, then it is always
possible to find a differential form
of {\it rank three}, satisfying Maxwell-type equations, such that, in a
coordinate basis
$$
F_{\mu\nu\rho}(X(\xi))=
{\rm const.}\times{{\dot X}_{\mu\nu\rho}\over\sqrt{-{1\over3!}
{\dot X}_{\alpha\beta\gamma}{\dot X}^{\alpha\beta\gamma}}}
$$
where ${\dot X}_{\mu\nu\rho}$ represents the tangent three-vector to the
membrane
world-track.The converse proposition is also  true.

In the first part of the paper, we show that a relativistic membrane,
regarded as a mechanical system,
admits a Hamilton-Jacobi formulation in which the H-J function describing
a family of classical membrane histories is given by
$\displaystyle{F=dB=dS^1\wedge dS^2\wedge dS^3}$ where the three scalar
functions $S^i(x)$ are the Clebsch potentials.

In the second part of the paper, we introduce a {\it new} lagrangian
of the Kalb-Ramond type which provides a {\it first order}
formulation for both open and closed membranes. The advantage of the
lagrangian approach is that it shows explicitly the correspondence
between the gauge formulation of the membrane in terms of the Kalb-
Ramond potential and the geometric formulation of the membrane in terms
of the mechanical coordinates $X^\mu(\xi)$. Finally, for completeness, we
show that such a correspondence can be established in the very general case
of a p-brane coupled to gravity in a spacetime of arbitrary dimensionality.
\bigskip

%\endpage
%%%%%%%%%%%%%%%%%%%%%%%%%%%%%%%%%%%%%%%%%%%%%%%%%%
\REF\Aa{%
Y.Nambu, Phys.Lett.{\bf 102B}, 149 (1981); {\bf 92B}, 327 (1980)
}%
\REF\Ah{%
H.A.Kastrup, Phys.Lett.{\bf 82B}, 237 (1979); Phys.Rep.{\bf 101}, 1 (1983)
\hfill\break
H.A.Kastrup, M.A.Rinke, Phys.Lett.{\bf 105B}, 191 (1981)\hfill\break
M.Rinke, Comm.Math.Phys.{\bf 73}, 265 (1980)
}%
\REF\Aj{%
T.Eguchi, Phys.Rev.Lett.{\bf 44}, 126 (1980)
}%
\REF\Ap{%
Y.Hosotani, Phys.Rev.Lett.{\bf 55}, 1719 (1985)
}%
\REF\Az{%
A.Aurilia, E.Spallucci
{\it ``~Hamilton-Jacobi formalism and the wave equation
for a relativistic membrane~''};
in preparation
}%
\REF\Al{%
A.Aurilia, E.Spallucci
{\it ``~The Role of Extended Objects in Particle
Physics and in Cosmology~''};
Proceedings of the Trieste Conference on
Super-Membranes and Physics in 2+1 dimensions, Trieste 17-21 July 1989;
eds.M.J.Duff, C.N.Pope, E.Sezgin; World Sci.
}%
\REF\Ai{%
A.Aurilia, F.Legovini, E.Spallucci, Phys.Lett.{\bf 264B}, 69 (1991)
}%
\REF\Av{%
A.Vilenkin, Phys.Rep.{\bf 121}, 263 (1985)\hfill\break
R.Basu, A.H.Guth, A.Vilenkin, Phys.Rev.{\bf D44}, 340 (1991)
}%
\REF\godb{%
C. Godbillon,
{\bf G\'eom\'etrie diff\'erentielle et m\'ecanique analytique};
Hermann, Paris 1969
}%
\REF\At{%
M.Rasetti, T.Regge, Physica {\bf 80A}, 217 (1975);\hfill\break
F.Lund, T.Regge, Phys.Rev.{\bf D14}, 1524 (1976)
}%
\REF\gold{%
H.Goldstein, {\bf Classical Mechanics},
Addison-Wesley, Reading, MA, (1980)
}%
\REF\york{%
J.D.Brown, J.W.York, Jr. Phys.Rev.{\bf D40}, 3312 (1989)
}%
\REF\segin{%
See, for example,
K.Kikkawa, Y.Masami, Progr.Theor.Phys.{\bf 76}, 1379, (1986);\hfill\break
E.Bergshoeff, E.Sezgin, P.K.Townsend,
Phys.Lett.{\bf 180B}, 370, (1986);\hfill\break
H.Luckock, I.Moss, Class.Quantum Grav.
{\bf 6}, 1993, (1989).\hfill\break
A much larger bibliography can be found in ref.\Al .
}%
\REF\hoso{%
C.Marshall, P.Ramond, Nucl.Phys.{\bf B85}, 375 (1975);
H. Choon-Lin, Y.Hosotani, Phys.Rev.Lett.{\bf 60}, 885, (1988);
H. Choon-Lin
{\it ``~Field Theory of Geometric p-branes~''};
UMN-TH 661/88, Minnesota Univ. Ph.D. Thesis.
}%
\REF\bryo{%
J.D.Brown, J.W.York, Jr. {\it Quasilocal energy and conserved charges
derived from the gravitational action}, IFP-423-UNC, (1992)
}%
\REF\Aw{%
A.Schild, Phys.Rev.{\bf D16}, 1722 (1977)
}%
\REF\Ab{%
A.Sugamoto, Nucl.Phys.{\bf B215}, 381 (1983)
}%
\REF\Ar{%
See Kastrup and Rinke, ref.2
}%
\REF\Ae{%
H.Rund,
{\bf The Hamilton-Jacobi theory in the calculus of variations};
Van Nostrand, London 1966.
}%
\REF\Ak{%
H.B.Nielsen, P.Olesen, Nucl.Phys.{\bf B57}, 367 (1973)
}%
\REF\Am{%
A.Aurilia, E.Spallucci, Phys.Lett.{\bf 282B}, 50 (1992)
}%
\REF\chodos{%
A.Chodos, C.B.Thorn, Nucl.Phys.{\bf B72}, 509 (1974)
}%
\REF\Au{%
H.A.Kastrup, Phys.Lett.{\bf 78B}, 39 (1978)
}%
\REF\As{%
A.Aurilia, E.Spallucci, Phys.Lett.{\bf 251B}, 39 (1990);\hfill\break
A.Aurilia, R.Balbinot, E.Spallucci, Phys.Lett.{\bf 262B}, 69 (1991)
}%
\REF\chs{C.Callan, J.Harvey, A.Strominger
\journal Nucl.Phys. &B 359 (91) 611
\journal Nucl.Phys. &B 367 (91) 60\hfill\break
 G.T.Horowitz, A.Strominger
\journal Nucl.Phys. &B 360 (91) 197\hfill\break
 M.J.Duff, J.X.Lu
\journal Phys. Lett. &B273 (91) 409
\journal Nucl.Phys. &B 390 (93) 276}
\REF\al{A.Aurilia, F.Legovini
\journal Phys. Lett. &B 67 (77) 509\hfill\break
A.Aurilia, D.Christodoulou, F.Legovini
\journal Phys. Lett. &B 73 (78) 429}
\REF\haw{S.W.Hawking
\journal Phys. Lett. &B134 (84) 403\hfill\break
A.Aurilia, A.Smailagic, E.Spallucci
\journal Class. Quantum Grav. &9 (92) 1883}
\REF\freund{ P.G.O.Freund, P.Oh, J.T.Wheeler
\journal Nucl.Phys. &B 246 (84) 371}
\REF\kuchar{%
K.Kuchar,
{\it ``~Canonical Methods of Quantization~''}, p.329, in
{\bf Quantum Gravity 2};
eds. C.J.Isham, R.Penrose, D.W.Sciama; Clarendon press, Oxford (1981)
}%

\refsend
%%%%%%%%%%%%%%%%%%%%%%%%%%%%%%%%%%%%%%%%%%%%%%%%%

\chapter{Introduction}
There is a remarkable but relatively unknown correspondence between
the dynamics of a relativistic string and a restricted class of
electromagnetic fields characterized by the condition ${}^* F^{\mu\nu}
F_{\mu\nu}=0$.
The key formula that links gauge fields to string coordinates is given
by $F^{\mu\nu}\left(x=X(\tau,\sigma)\right)=
({\rm const.})\,{\dot X}^{\mu\nu}/\sqrt{-{1\over
2}{\dot X}^{\alpha\beta}{\dot X}_{\alpha\beta}}$,
where ${\dot X}^{\mu\nu}$ represents the tangent element to the world
sheet of
the string\foot{If $x^\mu=X^\mu(\tau,\sigma)$ represents the embedding
of the string world-sheet in Minkowski spacetime, then
$$
{\dot X}^{\alpha\beta}\equiv {\partial(X^\alpha,X^\beta)\over\partial(\tau,
\sigma)}\ .
$$
}.
Even more remarkable is the fact that a relativistic string, regarded
as a mechanical system, admits a generalized Hamilton-Jacobi formulation
in
which the H-J function describing a family of classical string histories is
given by $H={1\over4}F_{\mu\nu}F^{\mu\nu}={\rm const.}$
These results where established many years ago by Nambu [\Aa],
Kastrup and Rinke [\Ah],
and lead one to speculate whether the correspondence between relativistic
strings and Maxwell fields represents a peculiar mathematical coincidence
or,
whether it represents a special case of a {\it general gauge field
representation
of geometric objects of any dimensionality}. With this question in mind,
the primary purpose of this paper is to show that a definite
relationship exists between relativistic membranes and gauge fields of
Kalb-Ramond type. We shall illustrate the precise sense of this
correspondence
by developing the Hamilton-Jacobi theory for a relativistic membrane
in a way which is analogous to the case of point particles and strings;
an obvious pattern emerges from this analysis and we show that our
results can be extended to a generic hypersurface, or p-brane, embedded
in a Riemannian manifold with an arbitrary number of dimensions.

Before we embark on a technical discussion of our results, let us deal
briefly with the obvious question: why should one undertake this program
of research in the first place?

There are many excellent reasons that one may advocate for this study:
for instance, Nambu's objective was to establish a definite relationship
between the string model as a phenomenological approach to the dynamics
of hadrons and the fundamental degrees of freedom of Quantum
Chromodynamics, with an eye on the long standing problem of quark
confinement; Kastrup and Rinke were more interested in the mathematical
aspects
of the correspondence between strings and gauge fields especially in
connection
with Carath\'eodory's H-J theory for fields, while Eguchi and
Hosotani [\Aj,\Ap] were exploring
an alternative route to the quantization of
relativistic strings. To our mind, all of the above arguments are equally
valid and in a subsequent paper we shall address the problem of deriving the
functional wave equation for membranes which reduces to the Hamilton-
Jacobi
equation in the classical limit [\Az].
However, more compelling from our vantage point is the fact that
relativistic
extended objects, either as
solutions of local quantum field theories or as fundamental geometric
structures in spacetime, are playing an increasingly important role
not only in particle physics but also in cosmology [\Al], especially
in connection with the physical processes leading to mass
generation [\Ai] and to the formation of structure in the early universe
[\Av].
Further
progress in this direction is possible only if we have a firm grasp of the
basic dynamical properties of extended objects.

The content of the paper can be summarized as follows:
Sect.2 provides the background of our work and is devoted to introduce
notation, conventions, and to define the geodesic field of a membrane.

In the first part of Sect.3 we derive the canonical
Hamilton-Jacobi equations for
a relativistic \m\ described by the Nambu-Goto action and discuss the
consequences of reparametrization invariance. In the second part of the
section
we introduce a non-canonical formalism, particularly suited to
treat reparametrization invariant theory. The main result of this section
is the set of {\it generalized Hamilton-Jacobi equations } for the membrane.

In Sect.4 we introduce a new Kalb-Ramond type field theory, both for open
and closed \ms , and discuss the
relationship between solutions of the generalized Maxwell equations and
solutions of the \m\ classical equations of motion.

In Sect. 5 we illustrate how the formalism works in four dimensions by
constructing the
explicit
gauge field representation of a spherical membrane.

In Sect. 6 we complete our discussion by extending the results of section 4
to the case of a p-brane embedded in a D-dimensional Riemannian manifold.
Interestingly enough, the gauge theory of a (D-1)-brane, or hyperbag, turns
out to be equivalent to the gauge formulation of the cosmological constant
in Einstein's equations.

Appendix A collects some useful geometrical definitions and
theorems, mainly concerning the rank and  class of a differential form
[\godb].

In Appendix B we give a brief derivation of the Hamilton-Jacobi theory for a
relativistic \m\ starting from a parametrized canonical hamiltonian.

Throughout the paper we shall express physical quantities in natural
units $\hbar=c=1$, and use metric signature $-+++$.

\chapter{Background and Definitions}

The crux of the correspondence between relativistic strings and
electromagnetic fields is the mathematical property that the Maxwell
field is a closed 2-form of {\it rank} 2. In Appendix A we define the
notion of `` rank '' of a differential form; roughly speaking, it means
that one can find a coordinate system in which
$F\equiv dA\equiv {1\over 2!}\ff\,
dx^\mu\wedge dx^\nu$ takes the form $F=dS^1\wedge dS^2$, or,
in component notation,
$\ff=\partial_{\,[\mu}A_{\nu]}=\partial_{\,[\mu}S^1\partial_{\nu]}S^2$.
Thus the gauge potential is $A_\mu=S^1\partial_\mu S^2=-S^2\partial_\mu
S^1$, one form being obtained from the other by the gauge transformation
$A\rightarrow A+d(S^1S^2)$. The two functions $S^1$ and $S^2$ are
interpreted as scalar potentials in the Hamilton-Jacobi theory of
relativistic strings [\Aa,\Ah].

Even at this early stage,
one cannot fail to observe that the number of independent scalar functions
$S^i$ is the same as the number of geometric dimensions of the world-
history
of the object in spacetime, i.e. 1-dimension for point particles
and 2-dimensions for strings [\Ah].
However, one may wonder why the H-J function for {\it strings} involves
the electromagnetic field which mediates the interaction between
{\it point-like} charges. The most naive answer that comes to our mind
is that a set of two point charges described by their world-lines define
the closed boundary of a string to which one associates a world-sheet
in spacetime; since $\ff\left(x=X(\tau,\sigma)\right)$ represents the
field
of tangential planes to the extremal surface, one expects that locally,
i.e. in the neighborhood of a point on the world sheet,
$F^{\mu\nu}\sim~{\dot X}^{\mu\nu}$.

To complete our preliminary discussion we note that the case of a
relativistic
string is exceptional in one respect which is worth spelling out now:
with reference to the definitions given in Appendix A, in four
dimensions the electromagnetic field $F={1\over2}\ff dx^\mu\wedge
dx^\nu$,
is of {\it rank} 4 and the associated system $\Lambda^*(F)$ is
spanned by four 1-forms $dx^\mu$. The only associated vector field is
$X=0$.
As a matter of fact, the equation $v^\mu\ff=0$ admits only the trivial
solution
$X=0$ if\  ${\rm det}\ff=(\vec E\cdot\vec B)^2\ne 0$.
However, if\  ${\rm det}F_{\mu\nu}=0$ and $F_{\mu\nu}\ne 0$, then $F$ has
rank
two and its associated space $A(F)$ can be spanned by two vector fields
$
X_1=B^j\partial_j\ ;\quad X_2=\partial_0{\vec B}^2
-{1\over2}(\vec E\wedge\vec B)^j\partial_j\ .
$
Furthermore, if $F$ has rank 2, its class is also 2 since $dF=0$. Then
the differential system above is integrable and defines two
submanifolds $S^i(x)={\rm const.}$, $i=1,2$ of $M^4$ which can be regarded
as 2-dimensional Hamilton-Jacobi wave fronts associated with the motion
of a relativistic string. {\it The restriction ${}^*F^{\mu\nu}\ff=0$
reduces the rank of $F$ from} 4 {\it to} 2.

On the basis of the foregoing discussion one is led to expect
that the Hamilton-Jacobi theory of a {\it relativistic membrane}, whose
world-track ${\cal H}$ is a 3-dimensional timelike submanifold of $M^4$,
requires three scalar functions $S^i$, $(i=1,2,3)$, and involves a
generalized `` Maxwell field '' of Kalb-Ramond type
$$
\f=\partial_{\,[\lambda}B_{\mu\nu]}=\partial_{[\lambda}S^1\wedge
\partial_\mu S^2\wedge\partial_{\nu]}S^3\ .
$$
One further expects that, if $\xi^a$ $(a=0,1,2)$ are local lorentzian
coordinates
parametrizing $\cal H$, then $\F(x)\sim{\dot X}^{\lambda\mu\nu}$ for
$x=X(\xi^a)$,
where
$$
{\dot X}^{\lambda\mu\nu}\equiv{\partial
X^\lambda\over\partial\xi^0}\wedge
{\partial X^\mu\over\partial\xi^1}\wedge{\partial X^\nu\over\partial\xi^2}
=\delta^{[abc]}\partial_a X^\lambda\partial_b X^\mu\partial_c X^\nu
$$
represents the tangent 3-vector to the world-history of the bubble in
spacetime.

To show that this is indeed the case, start from the Nambu-Goto
lagrangian $\L=\L({\dot X})$ for a closed membrane
$$
\L=-\r\sqrt{-{1\over 3!}{\dot X}^{\lambda\mu\nu}{\dot X}_{\lambda\mu\nu}}
=-\r\sqrt{-{\dot X}^{(\lambda\mu\nu)}{\dot X}_{(\lambda\mu\nu)}}
\eqn\lagra
$$
where $\rho$ represents the surface tension and
${\dot X}_{(\lambda\mu\nu)}$ denotes the {\it restricted components} of
${\dot X}$, i.e. ${\dot X}_{\lambda\mu\nu}$ with $\lambda<\mu<\nu$.

The stationary action principle,
when applied to the membrane action\hfill\break
$S=\int d^3\xi\,\L$, leads to the classical equations of motion
$$
\delta^{[abc]}\left(
{\partial\over\partial\xi^a}{\partial\L\over\partial{\dot
X}^{\lambda\mu\nu}}
\right)
{\partial X^{\mu}\over\partial \xi^b}{\partial X^{\nu}\over\partial \xi^c}
=0
\eqn\moto
$$
representing the conservation of the {\it volume canonical momentum}
$$
\Pi_{\mu\nu\rho}\equiv\partial\L/\partial{\dot X}^{\mu\nu\rho}=
\r{{\dot X}_{\lambda\mu\nu}\over\sqrt{-{1\over 3!}{\dot
X}_{\alpha\beta\gamma}\,
{\dot X}^{\alpha\beta\gamma}}}
\eqn\pigreco
$$
along the membrane history. For an open membrane one would find,
in addition,
the constraint that $\Pi_{\mu\nu\rho}$ must vanish along the boundary,
that is $\Pi_{\mu\nu\rho}{\dot X}^{\nu\rho}=0$, where
${\dot X}^{\nu\rho}$ represents
the tangent element to the boundary itself.

Equation \moto\ can also be written in the more conventional form
$$
{\partial\over\partial\xi^a}{\partial\L\over\partial(\partial X^\mu
/\partial\xi^a)}=0
\eqn\motob
$$
since
$$\eqalign{
{\partial\over\partial\xi^a}{\partial\L\over\partial(\partial X^\lambda
/\partial\xi^a)}&=
{\partial\over\partial\xi^a}\Bigl[{\partial\L\over\partial{\dot
X}^{\sigma\mu\nu
}}
{\partial{\dot X}^{\sigma\mu\nu}\over\partial X^\lambda
/\partial\xi^a}\Bigr]\cr
&={1\over2}
{\partial\over\partial\xi^a}\Bigl[\P\delta^{[abc]}
{\partial X^{\mu}\over\partial\xi^b}{\partial X^{\nu}\over\partial\xi^c}
\Bigr]=
{1\over2}\moto\ .\cr}
\eqn\motoc
$$
{}From eq.\lagra\ it is immediate to see that the volume
conjugate momentum satisfies the following relations
$$
\P(\xi)=-\r^2{{\dot X}_{\lambda\mu\nu}\over\L}\,,
\eqn\impa
$$
so that
$$
{1\over3!}\P\,{\dot X}^{\lambda\mu\nu}=
{\r^2\over3!}\,{{\dot X}_{\lambda\mu\nu}\over\L}{\dot
X}^{\lambda\mu\nu}=\L\ .
\eqn\impb
$$
The volume momentum differs from zero only along the world history of the
extended object, in the same way as the ordinary linear momentum, say
$P_\mu$,
is non-vanishing only along the world-line of a point particle. However,
suppose it is possible to define in a non-trivial way\foot{ The trivial
extension is
$$
\P(x)=\cases{\P(\xi), &for $x=X(\xi)$;\cr  0, &for $x\ne X(\xi)$.\cr}
$$
In this case, the field $\P(x)$ is nothing but the \m\ current
$J^{\mu\nu\rho}(x)$ except for a proportionality constant. However, this
is not the correct way to extend $\P(\xi)$ in order to formulate a {\it gauge
theory} for the \m\ (~see below and the discussion at the beginning
of Sect.4). }
a smooth field
$\Pi_{\mu\nu\rho}(x)$ over the spacetime manifold, which coincides with
$\P(\xi)$ along $x^\mu=X^\mu(\xi)$. Then, we can introduce the 3-form
$$
\Omega(x)={1\over3!}\,\P(x)\, dx^{\lambda}\wedge dx^{\mu}\wedge dx^{\nu}\
\eqn\formax
$$
When evaluated on the membrane, that is for $x=X(\xi)$, $\Omega$
is nothing but the action element:
$$
\Omega\left(x=X(\xi)\right)=
\P(\xi)\,{\dot X}^{(\lambda\mu\nu)}\,d\xi^0\wedge d\xi^1\wedge d\xi^2
=\L\,d\xi^0\wedge d\xi^1\wedge d\xi^2\ .
\eqn\formaxi
$$
The introduction of the field $\P(x)$ as the spacetime counterpart of
$\P(\xi)$
is not a trivial operation. The main difficulty is that there is a basic
ambiguity,in the
sense that many different $\Pi(x)$ can match $\Pi(\xi)$ along the \m\
world-track, so that the correspondence between $\Pi(x)$ and $\Pi(\xi)$ is
not necessarily one-to-one, thus causing integrability problems. However, in
analogy to the string case, there
is a particular {\it canonical extension} of $\P(\xi)$ which leads to a
gauge description of the \m\ in terms of {\it.three scalar potentials}. Such a
canonical extension hinges on the
properties of a special geometric
field which describes the tangent element to the \m\ world-surface.
A {\it slope field} $(x,\Phi(x))$ for a family of
extremals of the action $S$, is a 3-vector
$\Phi^{(\lambda\mu\nu)}(x)$ such that:
$$
{\dot X}^{(\lambda\mu\nu)}(\xi)=\Phi^{(\lambda\mu\nu)}\left(X(\xi)\right)\
\eqn\campo
$$
The slope field is a generalization of the velocity field: it is a field
in the sense that it is defined at every point
in spacetime, but when evaluated along
the history of the object ( point-particle, string, membrane,
\dots), it gives the corresponding {\it tangent element}. A non-relativistic
example of slope-field which has inspired us,
is found in the formulation of the dynamics of vortices
in a super-fluid [\At]. A {\it vortex} can be described as a closed
curve $\Gamma$ in $R^3$, parametrized by three functions
$x^i=X^i(\sigma)$;
the vortex moves in a fluid which is described by a divergenceless
velocity field $v^i=v^i(t,x^k)$. The dynamical input of the model is the
requirement that at any given point the vortex velocity coincides with the
fluid velocity
$$
{dx^i\over dt}=v^i\Big\vert_{x=X(\sigma)}\ .
$$
Then we can establish the following correspondence:
$$
\eqalign{
&\hbox{fluid velocity field}\Longleftrightarrow\hbox{slope field}\cr
&\hbox{vortex velocity}\Longleftrightarrow\hbox{tangent element}\ .\cr
}
$$
In terms of this geometric field the volume momentum can be extended
as follows
$$
\P(\xi)\longrightarrow \P(x,\Phi(x))\equiv
\r{\Phi_{\lambda\mu\nu}(x)\over\sqrt{-{1\over
3!}\Phi_{\alpha\beta\gamma}(x)
\Phi^{\alpha\beta\gamma}(x)}}\ .
\eqn\estens
$$
Then,it follows from the definition \estens, that
$\dis{\Pi\left(x=X,\Phi=\dot X\right)=\P(\xi)}$.

The slope field is said to be {\it geodesic} with respect to the Lagrangian
$\cal L$ if the form  $\Omega(x,{\dot X}=\Phi)$ is  closed:
$$
d\,\Omega(x)\equiv d\,[\,{1\over3!}\P(x)\,
dx^{\lambda}\wedge dx^{\mu}\wedge dx^{\nu}\,]=0\ .
\eqn\chiusa
$$
{\it Since $\Omega$ is a 3-form in four dimensions, and it is closed, then
its rank and class are three. Therefore, locally,
$\Omega$ can be written in terms of three exact differentials}
$$
\Omega(x)=dS^1(x)\wedge dS^2(x)\wedge dS^3(x)\ ,
\eqn\tred
$$
with
$$
\P(x)=S_{(\lambda\mu\nu)}(x)=
{\partial(S^1,S^2,S^3)\over
\partial(x^{\lambda},x^{\mu},x^{\nu})}\ .
\eqn\clebsh
$$
The functions $S^i(x)$ are the Clebsch potentials [\Aa,\Ah].

This property represents an essential difference with respect to
the string case: as mentioned earlier, a string can be described in terms of
a rank-2,
Maxwell two-form $F_{\mu\nu}$; but in four dimensions a two-form has
generally
rank-4. Thus, one imposes Pl\"ucker's conditions [\Ah]:
$\epsilon^{\mu\nu\rho\sigma}F_{\mu\nu}F_{\rho\sigma}=0$,
to reduce the rank of
the form to two.
In this sense the description of the membrane in terms
of a 3-form is more natural than the description of a string
in terms of a 2-form, since no extra conditions are required.

Finally, by definition the membrane is {\it imbedded }
in the geodesic field
$(x,\Phi(x))$ if there exists a function $\gamma(\xi)>0$ such that
$$
{\dot X}^{\lambda\mu\nu}(\xi)=\gamma(\xi)S^{\lambda\mu\nu}(X(\xi))\ .
\eqn\geod
$$
Then, we call $(x,\Phi(x))$ a {\it geodesic field for} $X(\xi)$, and the
field $\P(x)$ represents the desired canonical extension of $\P(\xi)$.

{\it
In what follows we show that the geodesic field of a membrane satisfies
generalized Maxwell equations, and that solving these equations is
equivalent
to solving the Nambu-Goto equation of motion $(2.2)$}.

\chapter{Hamilton-Jacobi Formalism}

In the first part of this section we shall outline the H-J description
of the classical dynamics of a relativistic \m. In the second part
we shall rephrase this approach in terms of Carath\'eory's formulation
of the H-J theory. The merit of the latter formulation is that it brings out
the explicit correspondence
between the geometric description and the gauge field description of
extended
relativistic objects.
\medskip
\underbar{\it Canonical H-J formulation}

The action for a closed \m\ in Minkowski spacetime is\foot{ In order to
avoid technical complications
we shall restrict our discussion to the case of closed \ms. This means that
the only boundary of the world-track swept in spacetime by the extended
object is represented by the initial and final \m\ configurations. In the
case of an open \m , an additional boundary contribution must be taken into
account, but this extra contribution does not affect the conclusions of this
section.}
$$
S=\int_D d^3\xi\,{\cal L}\ ,
\eqn\nga
$$
where the domain $D$ in parameter space is a three-surface
having the initial and final \m\ configurations as its only (~spacelike~)
boundary: $\dis{\partial D=\Sigma_{1)}\cup \Sigma_{2)}}$.

If
$\dis{x^\mu=X^\mu(\xi^k)}$ is the embedding of the \m\ world-track in
Minkowski
spacetime and $\dis{\xi_{i)}{}^k=\xi^k(s^m)\ ,\quad m=1,2}$ represent
the two components of the boundary, $i)=1),2)$,
in parameter space, then
$$\eqalign{
x^\mu&=X^\mu(\xi)\Big\vert_{\Sigma^1}\equiv X_{1)}{}^\mu(s^m)\ ,\cr
x^\mu&=X^\mu(\xi)\Big\vert_{\Sigma^2}\equiv X_{2)}{}^\mu(s^m)\ ,\cr
}\eqn\embedds
$$
are the corresponding embeddings of the initial, and final \m.

The action \nga\ is invariant under boundary preserving reparametrizations
$$\eqalign{
&X^\mu(\xi)\rightarrow \bar X^\mu(\sigma)=X^\mu\left(\sigma(\xi)\right)\
,\cr
&X^\mu(s^m)\Big\vert_{\Sigma^{1,2}}\rightarrow \bar X^\mu(\sigma(s^m))
\Big\vert_{\Sigma^{1,2}}=
X^\mu(s^m)\Big\vert_{\Sigma^{1,2}}\ .\cr}
\eqn\rep
$$

The volume momentum \impa implies the constraint
$$
H_0={1\over 3!}\p\P+\r^2=0\ .
\eqn\vincolo
$$
On the other hand, the canonical hamiltonian vanishes identically on account
of the
homogeneity in ${\dot X}^{\mu\nu\rho}$ of \lagra which, in turn, stems from
the requirement of reparametrization invariance:
$$
H_{\rm c}={1\over 3!}\P{\dot X}^{\lambda\mu\nu}-{\cal L}\equiv 0\ .
\eqn\hc
$$
Therefore, the equation of motion can be obtained by extremizing
the Hamilton-Jacobi action\foot{It is worthwhile to remark that
the least action principle in the Jacobi formulation describes
a system at ``~fixed energy~'', while the physical time lapse between the
initial and final configuration is free [\gold]. In our case, ``~the constant
value of the energy~'' is $\r^2$ and the physical time is replaced by the
proper volume of the \m\ world-tube. The Jacobi approach to General
Relativity
produces similar interesting results. The fixed value of the energy is, in
this case, the value of the cosmological constant [\york], which can be
viewed as the surface tension of the cosmic vacuum. } (~see Appendix B~)
$$
S[X(\xi),\,\Pi(\xi);\,N(\xi)]=\int_D d^3\xi\left({1\over 3!}\P
{\dot X}^{\lambda\mu\nu}-N H_0\right)
\eqn\hamact
$$
under variation of $X(\xi),\,\Pi(\xi)$ and $N(\xi)$ with the boundary
conditions
\embedds. $N(\xi)$ is a Lagrange multiplier enforcing the constraint
\vincolo, so that, varying S, one obtains

$$\eqalign{
\delta S=\int_Dd^3\sigma\Bigl[&
{1\over 3!}\delta\P{\dot X}^{\lambda\mu\nu}+{1\over 3!}\P
\delta{\dot X}^{\lambda\mu\nu}-{2\over 3!}
N\delta\p\P\cr
-&\delta N\left({1\over 3!}\p\P+\r^2\right)\Bigr]\ .\cr}
\eqn\primav
$$
Integration by parts enables us to isolate the boundary terms
$$
\eqalign{
\delta S=\int_Dd^3\sigma\Bigl[&
{1\over 3!}\delta\P\left({\dot X}^{\lambda\mu\nu}-2N
\p\right)-{1\over 2}\delta^{[abc]}\partial_a\p\partial_b X^\nu
\partial_c X^\rho\delta X^\mu\cr
-&\delta N H_0\Bigr]\cr
+\int_Dd^3\sigma &{1\over 3!}\delta^{[abc]}\partial_a\left(
\p\partial_b X^\nu\partial_c X^\rho\delta X^\mu\right)\cr}
\eqn\secondv
$$
and the resulting equations of motion are
$$
{\delta S\over\delta\p}=0:\quad {\dot X}_{\lambda\mu\nu}-2N \P=0\ ,
\eqn\uno
$$
$$
{\delta S\over\delta X^\mu}=0:\quad
\delta^{[abc]}\partial_a\P\partial_b X^\mu
\partial_c X^\nu=0\ ,
\eqn\due
$$
$$
{\delta S\over\delta N}=0:\quad {1\over 3!}\p\P+\r^2=0\ .
\eqn\tre
$$

On the basis of equations\uno\ and \tre\ we define
$$\eqalign{
&\Pcl={1\over 2\ncl}{\dot X}^{\mu\nu\rho}\ ,\cr
&\ncl={1\over 2\rho}\sqrt{-{1\over 3!}{\dot X}^{\mu\nu\rho}
{\dot X}_{\mu\nu\rho}}\ .\cr}
\eqn\csol
$$
Then, by inserting $\ncl$ in $\Pcl$ we obtain the volume momentum \impa
corresponding to a classical solution of eq.\due.

The purpose of the above manipulations is to lay the foundations of a future
theory of quantum membranes in the form of a semiclassical functional
wave equation. As a matter of fact, in the existing literature
there are two different attitudes towards the quantization of
extended objects. The first is to proceed as closely as possible to
the quantization of point-like particles. In modern language, that means
to study the spectrum of small oscillations around some suitable classical
solution playing the role of ground state of the system,
and then to interpret the excited states of the system as
``~particles~''[\segin].
The second approach is of a more geometrical nature and
consists in studying a suitable functional wave equation determining the
probability amplitude for a given
configuration of the \m\ [\hoso]
\foot{In this connection, it is worthwhile to note the close analogy of our
present discussion with the methods of covariant and canonical quantization
of gravity. In the first case the emphasis is on the concept of graviton
as the particle counterpart of the spacetime metric, and the main purpose is
to investigate graviton interactions with all the other elementary particles.
In the second case one studies the quantum geometry on a given spacelike
slice by introducing a ``~wave function of the universe~'' as a
solution of the Wheeler-De Witt functional wave equation. Both approaches
provide far reaching  insights into different aspects of gravity.}.

In the spirit of the second approach, we demand
that the variation of the membrane action be restricted to the set of
classical
solutions so that the terms in \secondv\ leading to the equations of motion
vanish together with the term which is proportional to $H_0$. Hence,
we find

$$\eqalign{
\delta S_{{\rm cl.}}&=
\int_Dd^3\sigma{1\over 3!}\delta^{[abc]}\partial_a\left(
\Pi_{{\rm cl.}\,\mu\nu\rho}\partial_b X^\nu\partial_c X^\rho\delta
X^\mu\right)
\cr
&={1\over 2}\int_{X_{1)}(s)}^{X_{2)}(s)}dx^\mu\wedge dx^\nu\pcl \delta
X^\rho(s)
\ .\cr}
\eqn\sbordo
$$
{}From here we obtain an expression for the classical volume momentum
evaluated at the final \m\ configuration
$$
{\delta S_{{\rm cl.}}\over \delta X_{2)}{}^\mu(s)}={1\over 2}
\pcl(s)X^{\nu\rho}(s)\ ,\quad X_{2)}{}^{\nu\rho}(s)\equiv
{\partial(X^\nu,X^\rho)\over\partial(s^1,s^2)}\Big\vert_{\Sigma_2}\ .
\eqn\hjuno
$$
Before proceeding, it is interesting to point out the close analogy between
our results \sbordo, \hjuno\ and the corresponding
Jacobi variation of the canonical
action for a non-relativistic particle as discussed, for instance, by Brown
and York [\bryo]. For a non relativistic particle, the reparametrized action
reads
$$
S=\int_{\tau_1}^{\tau_2}d\tau\left[p\dot x-{\dot t}H(x,p,t)\right]\ ,
\eqn\part
$$
where $\tau$ is a suitable parameter along the path of the system in
state space, and the coordinate time $t$ is now considered as a dynamical
variable $t=t(\tau)$. Varying the action (~by keeping fixed the end-points
$x_{1,2}=x(\tau_{1,2})$~) among the classical solutions of the equations
of motion we obtain
$$
\delta S_{\rm cl.}=
p_{\rm cl.}\delta x\Big\vert_{\tau_1}^{\tau_2}- H_{\rm cl.}(x,p,t)\delta t
\Big\vert_{\tau_1}^{\tau_2}\ .
\eqn\varcl
$$
The non-relativistic Hamilton-Jacobi equations follow immediately
from \varcl, e.g., we find the classical momentum and energy at the
final point as
$$
p_{\rm cl.}(x_2)={\partial S_{\rm cl.}\over\partial x_2}\ ,
\eqn\clmom
$$
$$
H_{\rm cl.}(x_2)=-{\partial S_{\rm cl.}\over\partial t_2}\ ,
\quad t_2=t(\tau_2)\ .
\eqn\ecl
$$
Eq.\hjuno\ is exactly the relativistic counterpart of \clmom. However, as
a consequence of reparametrization invariance, there is no equation which
corresponds to \ecl, i.e., there is no way to associate a canonical energy
to the relativistic \m.
Equation \hjuno\ is instrumental in interpreting the momentum constraint
as the Hamilton-Jacobi equation of a relativistic membrane. In fact, (~in
what follows we shall omit the label $2)$~)
$$\eqalign{
{\delta S_{{\rm cl.}}\over \delta X^\mu(s)}
{\delta S_{{\rm cl.}}\over \delta X_\mu(s)}&=
{1\over 4}
\pcl X^{\nu\rho}\Pi_{{\rm cl.}\,\nu'\rho'}^\mu X^{\nu'\rho'}\cr
&={1\over 3!}\Pcl\pcl {1\over 2}X^{\alpha\beta}X_{\alpha\beta}\ .\cr}
\eqn\semicl
$$
Then, by taking into account the definition \csol\ and setting
$\dis{\vert\vert X^{\alpha\beta}
\vert\vert\equiv {1\over 2}X^{\alpha\beta}X_{\alpha\beta}}$, we obtain the
Hamilton-Jacobi form of the momentum constraint [\kuchar]:
$$
{1\over \vert\vert X^{\alpha\beta}\vert\vert^2}
{\delta S_{{\rm cl.}}\over \delta X^\mu(s)}
{\delta S_{{\rm cl.}}\over \delta X_\mu(s)}=-\r^2\ .
\eqn\semikg
$$

Equation \semikg\ represents the desired result: it can be interpreted as the
semi-classical limit of a \m\ functional
relativistic wave equation; as such, it provides a suitable starting point
towards the
quantum mechanics of the \m\ considered as a geometrical extended object
rather than an ordinary matter field multiplet defined over the
three-dimensional world-track.

    The problem of deriving a functional wave equation corresponding to
\semikg, namely,
the Wheeler-De Witt equation for the \m\ functional,
will be investigated elsewhere [\Az]. Presently, we wish to introduce a
generalized Hamilton-Jacobi formalism and show how this novel approach
leads to a gauge theory of a relativistic \m.

\medskip
\underbar{\it Generalized Hamilton-Jacobi formulation}

We start by noting that the Nambu-Goto action \lagra\ of the membrane is
manifestly reparametrization invariant, and its
Lagrangian is a homogeneous function of degree 1 with respect to the
generalized
velocity ${\dot X}$. Then
$$
{\dot X}^{(\lambda\mu\nu)}{\partial\L\over \partial{\dot
X}^{(\lambda\mu\nu)}}=
\cal L\ ,
\eqn\omo
$$
and repeated application of $\partial/\partial{\dot X}^{(\lambda'\mu'\nu')}$
yields
the identity
$$
{\partial^2\L\over \partial{\dot X}^{(\alpha\beta\gamma)}
\partial{\dot X}^{(\lambda\mu\nu)}}{\dot X}^{(\lambda\mu\nu)}\equiv 0
\eqn\ident
$$
so that
$$
{\rm det}\left({\partial^2\L\over\partial{\dot X}^{(\lambda\mu\nu)}
\partial{\dot X}^{(\lambda'\mu'\nu')}}\right)=0\ .
\eqn\determ
$$
As a consequence, ${\dot X}$ cannot be expressed as a function of
(the coordinates and) the canonical momentum, and the
canonical Hamiltonian vanishes. Therefore, one has to introduce new
canonical
variables and the corresponding Hamiltonian.
One possibility is to extend Schild's formulation for strings [\Aw]
in which the action is not
reparametrization invariant; Schild's approach
was considered as a possible starting point for a canonical Hamilton-Jacobi
theory both for strings [\Ah,\Aj] and membranes [\Ab].
However,
this approach seems to be affected by integrability problems [\Ar].

We shall follow an alternative route: we maintain the Nambu-Goto action
but develop a new (non-canonical) formalism in which the key
ingredient is the
non-degenerate, {\it  generalized fundamental tensor} [\Ae]
$$
\G_{\alpha\beta\gamma,\lambda\mu\nu}\equiv-{1\over 2\r^2}\,
{\partial^2\L^2\over \partial{\dot X}^{(\alpha\beta\gamma)}
\partial{\dot X}^{(\lambda\mu\nu)}}
\eqn\funda
$$
and its inverse
$$
\G_{\lambda\mu\nu,\sigma\tau\eta}
\G^{\sigma\tau\eta,\alpha\beta\gamma}=
\delta_{[\lambda\mu\nu]}{}^{[\alpha\beta\gamma]}.
\eqn\inverso
$$
In this reformulation of the theory the role of canonical momentum
is assigned to the {\it new} quantity
$$
P_{\lambda\mu\nu}\,\equiv
-\r\,\G_{\lambda\mu\nu,\sigma\tau\eta}{\dot X}^{\sigma\tau\eta} .\
\eqn\mom
$$

Since the definition $(3.20,3.21)$ implies
$$
\G_{\alpha\beta\gamma,\lambda\mu\nu}=-{1\over\r^2}\left[
{\partial\L\over \partial{\dot X}^{(\alpha\beta\gamma)}}
{\partial\L\over \partial{\dot X}^{(\lambda\mu\nu)}}+
\L{\partial^2\L\over \partial{\dot X}^{(\alpha\beta\gamma)}
\partial{\dot X}^{(\lambda\mu\nu)}}\right]\ ,
\eqn\superg
$$
and in view of the homogeneity relations $(3.21,3.22)$,
the new momenta take on the following form
$$
P_{(\lambda\mu\nu)}={{\cal L}\over\r}
{\partial\L\over \partial{\dot X}^{(\lambda\mu\nu)}}=
{1\over 2\r}{\partial\L^2\over \partial{\dot X}^{(\lambda\mu\nu)}}\ .
\eqn\pielle
$$
Thus, we can alternatively interpret $P_{(\lambda\mu\nu)}$ as the
{\it dynamical variable
conjugate to $2\r{\dot X}^{(\lambda\mu\nu)}$ with respect to the new
Lagrangian
$\L^2$}.
Moreover, equation \mom\ yields
$$
{\dot X}^{\alpha\beta\gamma}=-{1\over\r}
\G^{\alpha\beta\gamma,\lambda\mu\nu} P_{\lambda\mu\nu}
\eqn\xgpi
$$
and thanks to the homogeneity relations \omo,\ident, it is easy to
verify that a double multiplication of \funda\ by ${\dot X}$, yields
the lagrangian squared
$$
\L^2=-\r^2\G_{\alpha\beta\gamma,\lambda\mu\nu}{\dot
X}^{\alpha\beta\gamma}
{\dot X}^{\lambda\mu\nu}\ .
\eqn\lquadro
$$
Equations \pielle,\xgpi,\lquadro\ reflect the rationale of the above
procedure:
the formal analogy between \lquadro\ and the generally covariant lagrangian
for
a ``~point-like particle~'' of ``~mass~'' $2\r^2$, moving in a {\it
Riemannian super-space} endowed with a ``~metric~''
$\G_{\alpha\beta\gamma,\lambda\mu\nu}$, and
``~four-velocity~'' ${\dot X}$, immediately suggests the definition of a
{\it Hamiltonian function } $H(P)$ defined through the Legendre transform
$$
H(P)^2\equiv P_{\lambda\mu\nu}\,\left(2\r{\dot
X}^{\lambda\mu\nu}\right)-\L^2
=-\G_{\alpha\beta\gamma,\lambda\mu\nu}
P^{\alpha\beta\gamma}P^{\lambda\mu\nu}\ .
\eqn\hquadro
$$
{}From the definitions \mom,\hquadro\ one can verify the following
reciprocity relations
$$
\G_{\alpha\beta\gamma,\lambda\mu\nu}\equiv-{1\over 2}\,
{\partial^2\L^2\over \partial P^{(\alpha\beta\gamma)}
\partial P^{(\lambda\mu\nu)}}\ ;
\eqn \rec
$$
$$
\r{\dot X}^{\alpha\beta\gamma}=H{\partial H\over\partial
P_{\alpha\beta\gamma} }
\;\eqn \xhpi
$$
$$
H^2\left(P={\L\over\r}\P\right)=-\r^2
\G_{\alpha\beta\gamma,\lambda\mu\nu}{\dot X}^{\alpha\beta\gamma}
{\dot X}^{\lambda\mu\nu}=\L^2\ .
\eqn \hl
$$
Finally, from eqs.$(\rec,\xhpi,\hl)$, in view of the symmetrical reciprocity
between $\L$ and $H$, we assume that the sign of $\L$ and $H$ must
coincide:
$$
H(P)=\L({\dot X})\ .
\eqn \hplx
$$
Now we have at our disposal
a suitable Hamiltonian formalism characterized by the momentum
$P_{\lambda\mu\nu}$
as defined in \mom.
Note that $H$ is a positive-homogeneous function of the first order and
of degree 1 with respect to $P_{\lambda\mu\nu}$,
$$
H(\kappa P)=\kappa H(P)\ .
\eqn\scala
$$
The corresponding Hamilton-Jacobi equations are:
$$\eqalign{
&{\partial H\over\partial
P_{(\lambda\mu\nu)}}={{\dot X}^{(\lambda\mu\nu)}\over\L}
\cr
&{\partial H\over\partial X^{\alpha}}=-{\partial \L\over \partial X^{\alpha}}
=0\ .\cr}
\eqn\hjeq
$$
Finally, from these equations and the homogeneity condition $(\scala)$
we obtain the
{\it generalized Hamilton-Jacobi equation}
$$
H\left(X,P={\L\over\r}{\partial\L\over\partial{\dot X}}\right)=
\L(X,{\dot X})\ \Rightarrow\  H(X,{\dot X})=\r
\eqn\hjgen
$$
in a form which is applicable to our specific problem.
For the Nambu-Goto system under consideration, the generalized Hamilton-
Jacobi equation becomes a ``~field equation~'' under the canonical extension
described in Sect.2
$$
H\left(x,S_{(\lambda\mu\nu)}(x)\right)=\r\
\eqn\hjinx
$$
i.e.
$$
[-S_{(\lambda\mu\nu)}S^{(\lambda\mu\nu)}]^{1/2}=\r
\eqn \squadro
$$
where
$$S_{\lambda\mu\nu}=\partial_{[\lambda}S^1\partial_{\mu}S^2
\partial_{\nu]}S^3\ .
\eqn \tres
$$
Thus, the essence of the above formalism is to select as the `` hamiltonian ''
the square root of the momentum-squared; then, the Hamilton-Jacobi
eq.\hjgen\
represents the square root of a generalized mass-shell condition for
$\P$.

It is perhaps instructive, at this point, to check that this formalism
reproduces well known results in the simple case of a point-like particle:
the relativistic Lagrangian for a point-like particle of mass $m$, moving
along a world-line $x^\mu=X^\mu(\tau)$, is
$$
L=-m\sqrt{-{\dot X}^\mu{\dot X}_\mu}\ ,\quad
{\dot X}^\mu\equiv{d X^\mu\over d\tau}\ .
\eqn\punto
$$
In this case the fundamental tensor \superg\ reduces to the
usual metric tensor:
$$
{\cal G}_{\mu\nu}\equiv-
{1\over 2m^2}{\partial^2
L^2\over\partial{\dot X}^\mu\partial{\dot X}^\nu}=\eta_{\mu\nu}\ ,
\eqn\metrica
$$
and the momentum is simply
$$
P_\mu\equiv-m\eta_{\mu\nu}{\dot X}^\nu={1\over 2m}{\partial
L^2\over\partial{\dot X}^\mu}
={L\over m}{\partial L\over\partial{\dot X}^\mu}\ .
\eqn\linmom
$$
Now, the square of
the Hamiltonian defined according to the Legendre transform \hquadro\
is
$$
H^2=(2m{\dot X}^\mu)P_\mu-L^2=-P_\mu P^\mu\ ,
\eqn\equadro
$$
and, with the choice $H(P)=L({\dot X})$, the H-J equation \hjgen\ reads
$$
H=m
\eqn\hm
$$
which is nothing but the square-root of the mass-shell condition:
$\displaystyle{-P_\mu P^\mu=m^2}$.
The corresponding ``~field theory~'' can be expressed in terms of the
slope field $U^\mu(x)$:
$$
U^\mu\left(x=X(\tau)\right)={P^\mu\over\sqrt{-P_\nu P^\nu}}=
-{\dot X^\mu\over\sqrt{-{\dot X}^\nu {\dot X}_\nu}}\ ,
\eqn\vel
$$
which coincides, along the particle world-line, with the unit norm 4-
velocity.
Thus, $U^\mu(x)$ can be interpreted as a relativistic velocity field,
consistently
with our previous remarks in Sect.2.
The action element \formaxi\ is now the 1-form
$$
\Omega(x)=m\, U_\mu(x) dx^\mu\ .
\eqn\densaz
$$
If $\Omega$ is closed, then $U^\mu(x)$ is the geodesic field embedding the
point particle:
$$
d\Omega=0\Rightarrow \partial_{\,[\mu} U_{\nu]}=0\Rightarrow
U_\mu(x)={1\over m}\,\partial_\mu S(x)\ .
\eqn\desse
$$
As expected, {\it only one Clebsch potential is required for a pointlike
object}.
The Hamilton-Jacobi equation \hm\ turns into the field equation
$$
\partial_\mu S \partial^\mu S +m^2=0\ ,
\eqn\wkg
$$
{\it which is just the WKB approximation to the quantum Klein-Gordon
equation
for a particle of mass $m$ described by the scalar field $\phi(x)$.} Indeed by
setting $\dis{\phi(x)\prop\exp {i\over\hbar}S(x)}$ in the Klein-Gordon
equation
one obtains eq.\wkg\ in the leading order in $1/\hbar$.

The application of this formalism to the case of a spherical bubble
will be discussed in section 5.

Returning to our general discussion,
the link between this generalized H-J formulation and the Lagrangian
description is provided by the following theorem:
\smallskip
{\it
If $S^j(x)$, $j=1,2,3$ is a solution of the H-J eq.\hjinx\
and $x^\mu=X^\mu(\xi)$ is a solution of $\P=S_{(\lambda\mu\nu)}(X(\xi))$,
then
$x^\mu=X^\mu(\xi)$ solves the Lagrange equations obtained from \lagra .
}
\smallskip
The general proof is given in ref.[\Ah] (Phys. Rep. sec.7.3 ).
In our case it implies that if $\P(\xi)=S_{\lambda\mu\nu}
\left(x=X(\xi)\right)$, then $x=X(\xi)$
is a solution of the classical equations of motion of the membrane.
Let us verify that this is indeed the case: consider the matrix
$\Delta_a{}^b\equiv\partial_a S^b\equiv u_a{}^\mu\partial_\mu S^b$, where
$u_a{}^\mu=\partial X^\mu/\partial\xi^a$. Now,
$$
{\partial\L\over\partial u_a{}^\mu}=
{\partial\L\over\partial{\dot X}^{(\lambda\nu\rho)}}
{\partial{\dot X}^{(\lambda\nu\rho)}\over\partial u_a{}^\mu}=
S_{(\lambda\nu\rho)}{\partial{\dot X}^{(\lambda\nu\rho)}\over\partial
u_a{}^\mu}
\ .\eqn \uno
$$
The antisymmetry of the determinant allows us to write
$$
{\rm det}(\Delta_a{}^b)=
S_{(\lambda\mu\nu)}(x){\dot X}^{(\lambda\mu\nu)}(u)\ .
\eqn\due
$$
Therefore, eq.$(\uno)$ can be written as
$$
{\partial\L\over\partial u_a{}^\mu}={\partial{\rm det}(\Delta_k{}^l)
\over\partial u_a{}^\mu}
={\bar\Delta}_b{}^c{\partial\over\partial u_a{}^\mu}\Delta_c{}^b
={\bar\Delta}_b{}^c{\partial\over\partial u_a{}^\mu}[(\partial_\nu S^b)
u_c{}^\nu]
=\Delta_b{}^a\partial_\mu S^b\ . \eqn\tre
$$
The bar refers to the co-factor
($\Delta{\bar\Delta}={\bar\Delta }\Delta={\rm det}\Delta$).
Since the derivative of a co-factor is given by
$$
d{\bar\Delta}_b{}^a=
({\rm det}\Delta)^{-1}
[{\bar\Delta}_b{}^a{\bar\Delta}_d{}^c-{\bar\Delta}_d{}^a{\bar\Delta}_b{}^c]
d\Delta_c{}^d
\eqn\quattro
$$
we have
$$
{\partial{\bar\Delta}_b{}^a\over\partial\xi^a}=
({\rm det}\Delta)^{-1}[{\bar\Delta}_b{}^a{\bar\Delta}_d{}^c-
{\bar\Delta}_d{}^a{\bar\Delta}_b{}^c]
{\partial^2S^d\over \partial\xi^a \partial\xi^c}=0\ .
\eqn\cinque
$$
Therefore
$$
{\partial\over\partial \xi^a}{\partial\L\over\partial u_a{}^\mu}=
{\bar\Delta}_b{}^a(\partial_\rho\partial_\mu S^b)u_a{}^\rho\ .
\eqn\sei
$$
On the other hand, we have
$$
{\partial\over\partial(\partial_\mu S^a)}\Delta=
{\dot X}^{(\lambda\nu\rho)}
{\partial S_{(\lambda\nu\rho)}\over\partial(\partial_\mu S^a)}\
,\eqn\sette
$$
and also
$$
{\partial\over\partial(\partial_\mu S^a)}\Delta=
{\bar\Delta}_c{}^b{\partial\over\partial(\partial_\mu S^a)}
[(\partial_\rho S^c)u_b{}^\rho]={\bar\Delta}_c{}^b u_b{}^\mu \ .
\eqn\otto
$$
Comparing \sette\ and \otto\ we find
$$
{\bar\Delta}_c{}^b u_b{}^\mu=
{\dot X}^{(\lambda\nu\rho)}
{\partial S_{(\lambda\nu\rho)}\over\partial(\partial_\mu S^a)}\ .\eqn\nove
$$
Thus we conclude that
$$\eqalign{
{\partial\over\partial \xi^a}{\partial\L\over\partial u_a{}^\mu}&=
(\partial_\rho\partial_\mu S^b){\bar\Delta}_b{}^c u_c{}^\mu
={\dot X}^{(\alpha\beta\gamma)}\partial_\mu S_{(\alpha\beta\gamma)}\cr
&={\dot X}^{(\alpha\beta\gamma)}\partial_\mu
{\partial\L\over\partial{\dot X}^{(\alpha\beta\gamma)}}=\partial_\mu\L=0
\cr}
\eqn\dieci
$$
that is, the Lagrange equations are satisfied, and our assertion is proved.
We shall make use of this theorem in the next section to establish
the relation between H-J functions for the membrane and  generalized
rank-3 Maxwell fields.

\def\sc{{\cal S}^{\rm closed}}
\def\so{{\cal S}^{\rm open}}
\def\w{W_{\mu\nu\rho}}
\def\W{W^{\mu\nu\rho}}
\def\b{B_{\mu\nu}}

\def\j{J_{\mu\nu\rho}}
\def\J{J^{\mu\nu\rho}}
\def\ki{\Phi_{\mu\nu\rho}}
\def\Gi{\Phi^{\mu\nu\rho}}
\def\st{{1\over 2\pi\alpha'}}
\def\cs{J_{(1)}}
\def\Z{\dot Z}

\chapter{The Membrane Maxwell Field}

Having discussed the mathematical background underlying the gauge
description of the relativistic \m, we wish, now, to translate the
results of the previous sections in a language which is more familiar
to the theoretical physicists, i.e. lagrangian field theory. More
precisely, we wish to show that {\it the geodesic field can be constructed
as a solution of a suitable
set of field equations.} However, in implementing this idea, it is important
to distinguish carefully between the slope field and the geodesic field in the
starting lagrangian, since only {\it on-shell} one finds a relation between
the two fields. As a matter of fact, the slope field and the geodesic field
play altogether different roles: the slope field encodes the information
about the \m\ geometric structure,
while the geodesic field describes the gauge properties of the membrane
and, by definition,
is the covariant curl of a gauge potential. With this distinction in mind, the
following approach can be interpreted as a
{\it first order} formulation of the usual geometric theory of \ms.

    Of central importance to this {\it new} approach are certain properties of
the
\m\ current which we list below. Irrespective of
whether the \m\ is closed or open, the associated current
$$
\J=\int_{D}d^3\xi\,\delta^{4)}\left(x-X(\xi)\right){\dot
X}^{\mu\nu\rho}
\eqn\corrente
$$
has the following properties:

\item{a)} when computed along the \m\ world-track, $\J$ is proportional
to the unit norm tangent element, that is

$\dis{\J(x=X)={\rm const.}{\dot X}^{\mu\nu\rho}/\sqrt{-{1\over 3!}
X^{\alpha\beta\gamma}X_{\alpha\beta\gamma}} }$;
\item{b)} the dual current is proportional to the normal to the \m\
world-track, i.e., it is orthogonal to $\J$, i.e.
$\dis{\epsilon_{\alpha\mu\nu\rho}\J J^{\alpha\beta\gamma}=0}$.
\medskip
Furthermore,
\item{c)} If $\partial D=\emptyset$, that is, if the \m\ is spatially
closed and infinitely extended along the time-like direction, then
$\dis{\partial_\mu\J=0}$;
\item{d)} suppose the \m\ is open and its boundary is a closed string.
If we parametrize the timelike boundary of the \m,\hfill\break
$\partial D$ as $\dis{\xi=
\xi(s^i)\ ,\quad i=0,1}$ then, $\dis{x^\mu=X^\mu\left(\xi(s^i)\right)=
X^\mu(s^i)}$ describes the word-sheet of a closed string, and
$$
\partial_\mu\J(x)=J^{\nu\rho}(x)\ ,\qquad\partial_\nu J^{\nu\rho}=0\
,\qquad
J^{\nu\rho}\equiv \int_{\partial D}d^2 s^i\,
\delta^{4)}\left(x-X(s^i)\right){\dot X}^{\nu\rho}\ .
\eqn\scorr
$$
\medskip
Property a) does not follow automatically from the definition \corrente\
because
of the singularity that arises along the transverse direction in the
coincidence limit
in the argument of the delta-function
$$
\delta^{4)}\left(x-X(\xi)\right)={\delta^{3)}(\xi'-\xi)\over\sqrt{-{1\over 3!}
X^{\alpha\beta\gamma}X_{\alpha\beta\gamma}}}
\delta(x'_\perp-x_\perp)\ .
\eqn\dirdel
$$
This divergence stems from the use of the ``~thin wall~'' approximation and
can
be avoided by assigning a physical
width $a$ to the \m , or, in mathematical terms, by approximating
$\delta(x'_\perp-x_\perp)$ with a
gaussian of the same width. Then,
$\dis{\delta(x'_\perp\rightarrow x_\perp)\vert_{\rm reg}=1/4a\sqrt\pi}$,
and
$$\eqalign{
\J(x=X)&=\int_{D}d^3\xi'\,
\delta^{4)}\left(X(\xi-X(\xi')\right){\dot X}^{\mu\nu\rho}\cr
&={{\rm const.}\over a}\int_{D}d^3\xi'\,
\delta^{3)}(\xi-\xi'){{\dot X}^{\mu\nu\rho}\over\sqrt{-{1\over 3!}
X^{\alpha\beta\gamma}X_{\alpha\beta\gamma}}}\cr
&={\rm const.}'\Pi^{\mu\nu\rho}(\xi)\cr}\ .
\eqn\propa
$$
Property b) is straightforward:
$$\eqalign{
&\epsilon_{\alpha\mu\nu\rho}\J J^{\alpha\beta\gamma}=\cr
&\epsilon_{\alpha\mu\nu\rho}
\int_{D}d^3\xi\,d^3\xi'\,\delta^{4)}\left(x-X(\xi)\right)
\delta^{4)}\left(x-X(\xi')\right)
{\dot X}^{\mu\nu\rho}{\dot X}'{}_{\mu\nu\rho}=\cr
&\epsilon_{\alpha\mu\nu\rho}
\int_{D}d^3\xi\,d^3\xi'\,\delta^{4)}\left(x-X(\xi)\right)
\delta^{4)}\left(x-X(\xi')\right)\delta^{[abc]}\delta^{[a'b'c']}\times\cr
&\partial_{a'} X^\alpha\partial_a X^\mu\partial_b X^\nu\partial_c X^\rho
\partial_{b'} X^\beta\partial_{c'} X^\gamma\cr
&\equiv 0\cr}
\eqn\jjdual
$$
since there is no totally antisymmetric tensor in the $a'abc$ indices
in three dimensions.

With properties a) and b) in hands, it would be tempting to identify $\J(x)$
with
$\Phi^{\mu\nu\rho}(x)$. However, there is a substantial difference between
these
two entities, namely, $\J$ is a {\it distribution} with support
along the \m\ world-track, while $\Phi^{\mu\nu\rho}(x)$ is a smooth,
regular tensor field defined over the whole spacetime. Accordingly,
we suggest the following relation between them
$$
\J(x)=\Phi^{\mu\nu\rho}(x)
\int_{D}d^3\xi\,\delta^{4)}\left(x-X(\xi)\right)\ ,
\eqn\corrslope
$$
which is our own definition of $\Phi^{\mu\nu\rho}(x)$ in terms of $\J(x)$.

Property c) is valid for closed \ms, and can be proved as follows
$$\eqalign{
\partial_\mu\J(x)&=\int_{D}d^3\xi\,{\dot
X}^{\mu\nu\rho}\partial_\mu
\delta^{4)}\left(x-X(\xi)\right)\cr
&=-\int_{D}d^3\xi\,\delta^{[abc]}\partial_b X^\nu\partial_c X^\rho
\partial_a
\delta^{4)}\left(x-X(\xi)\right)\cr
&=-\int_{\partial D\equiv\emptyset}d\xi^b\wedge d\xi^c\,
\partial_b X^\nu\partial_c X^\rho
\delta^{4)}\left(x-X(\xi)\right)\cr
&\equiv 0\ .\cr}
\eqn\zero
$$
Finally, property d) is a special case of the geometric relation
$$
\partial_{\mu_1}J^{\mu_1\mu_2\dots\mu_p}_{(p)}(x)=
J^{\mu_2\dots\mu_p}_{(p-1)}(x)\ ,
\eqn\catene
$$
involving the set of five possible {\it p-chains}, or de Rham current
distributions, with support in Minkowski spacetime,
$$\eqalign{
&J_{(0)}(x)=\delta^{4)}\left(x-X(\xi)\right)\ ,\qquad (p=0\ ,\quad
\hbox{by definition})\cr
&J^{\mu_1\mu_2\dots\mu_p}_{(p)}(x)=
\int_{D}d^p\xi\,{\dot X}^{\mu_1\mu_2\dots\mu_p}
\delta^{4)}\left(x-X(\xi)\right)\ ,\quad(p\ge 1)\ .\cr}
\eqn\derham
$$
The divergence operation maps each $p$-chain into the $(p-1)$-chain
associated
with the boundary of the world-history of the extended object. If, as in
the case c), there is no boundary, then eqs (4.8,4.9) yield zero. Moreover,
repeated application of the divergence operator maps a p-chain to
zero because a boundary has no boundary.

%%%%%%%%%%%%%%%%%%%%%%%%%%%%%%%%%%%%%%%%%%%%%%
Now we are ready to introduce and discuss a non-linear Lagrangian for the
membrane
geodesic field, which is partly suggested by the physical interpretation
of membranes
as {\it extended solitons} of an underlying local field theory, and partly
by the string non-linear electrodynamics proposed
some years ago by Nielsen and Olesen [\Ak].

The two cases of open and closed membranes have to be discussed
separately.

\underbar{\it closed membrane}

Let us consider the following action
$$\eqalign{
&\sc=-g^2\int d^4x\,\sqrt{-{1\over 3!}\w\W}+
{1\over 3!}\int d^4x\,\W\partial_{\,[\mu}B_{\nu\rho]} \cr
&\f(x)\equiv\partial_{\,[\mu}B_{\nu\rho]}(x)\ , \cr}
\eqn\azionec
$$
where $\w(x)$ is a totally antisymmetric tensor and $g$ is a dimensional
constant. Physical dimensions are assigned as follows: $[\w]=[\f]=[g^2]=
{\rm M}^2$. The $B$-field which appear in \azionec\ as a lagrange multiplier
enforcing the transversality of $\W$, will be identified on-shell,
with the \m\ gauge potential.

Varying the action \azionec\ with respect to $\b$ and $\w$, we get the
following set of field equations
$$
\partial_\mu\W=0
\eqn\dew
$$
$$
g^2{\w\over\sqrt{-{1\over
3!}W_{\alpha\beta\gamma}W^{\alpha\beta\gamma}}}
+F_{\mu\nu\rho}=0\ .
\eqn\fmunu
$$
The {\it closed} membrane is represented by a special solution of \dew,
namely
$$\eqalign{
&\hat\W(x)=m\int_Dd^3\xi\,\delta^{4)}\left(x-
X(\xi)\right){\dot X}^{\mu\nu\rho}=
m\,J^{\mu\nu\rho}(x)\ ,\cr
&m={\rm const.}\ ,\quad [m]={\rm M}\ .\cr}
\eqn\sol
$$
The right hand side of \sol\ is, except for a multiplicative constant,
the current distribution associated with the three-dimensional manifold
representing the history of a closed \m. According to c)
$\J$ has {\it vanishing divergence}.
Equation \fmunu\ then gives
$$
{\hat F}_{\mu\nu\rho}=-g^2{\hat\w\over\sqrt{-{1\over 3!}\hat
W_{\alpha\beta\gamma}
\hat W^{\alpha\beta\gamma}}}=-g^2
{\j\over\sqrt{-{1\over 3!}J_{\alpha\beta\gamma}J^{\alpha\beta\gamma}}}
\eqn\soldue
$$
from which it follows that
$$
-{1\over 3!}{\hat F}_{\lambda\mu\nu}{\hat F}^{\lambda\mu\nu}=g^4\ .
\eqn\fquadro
$$
Therefore, $F={1\over3!}\f\,dx^\lambda\wedge dx^\mu\wedge dx^\nu$
(which is {\it closed} by definition)
is a 3-form which satisfies (on-shell) the generalized H-J equation
\fquadro.
In view of our definition \corrslope\ we can write eq.\sol\  as
$$
\hat\W(x)=m\,\Gi(x)\int_Dd^3\xi\, \delta^{4)}\left(x-X(\xi)\right)\ ,
\eqn\campo
$$
then
$$\eqalign{
\hat F^{\mu\nu\rho}&=-g^2{\Phi^{\mu\nu\rho}(x)
\int_Dd^3\xi \delta^{4)}\left(x-X(\xi)\right)
\over\sqrt{-{1\over 3!}\Phi_{\alpha\beta\gamma}(x)
\Phi^{\alpha\beta\gamma}(x)
\left(\int_Dd^3\xi \delta^{4)}\left(x-X(\xi)\right)\right)^2}}\cr
&=-g^2{\Phi^{\mu\nu\rho}(x)\over\sqrt{-{1\over
3!}\Phi_{\alpha\beta\gamma}(x)
\Phi^{\alpha\beta\gamma}(x)}}\ .
\cr}\eqn\fslope
$$
The net result of these manipulations is
that while $\hat\w(x)\sim \j(x)$ is a {\it singular} field having support
only
along the membrane history, $\hat F^{\mu\nu\rho}\sim
\Phi^{\mu\nu\rho}(x)$
is defined
{\it over the whole spacetime manifold. } However, when evaluated
on the membrane world-track, $F_{\mu\nu\rho}$ is proportional to the
volume
conjugate momentum. In fact
$$
{\hat F}_{\mu\nu\rho}\left(x=X(\xi)\right)=
-g^2{{\dot X}_{\mu\nu\rho}\over\sqrt{-{1\over 3!}{\dot
X}_{\alpha\beta\gamma}
{\dot X}^{\alpha\beta\gamma}}}\equiv {1\over m}\Pi_{\mu\nu\rho}\ .
\eqn\geod
$$
Conversely, eq.\geod\ defines the {\it canonical volume
field} $\Pi_{\mu\nu\rho}(x)$:
$$
\Pi_{\mu\nu\rho}(x)\equiv m\hat F_{\mu\nu\rho}(x)=
-\r{\ki(x)\over\sqrt{-{1\over 3!}\Phi_{\alpha\beta\gamma}
\Phi^{\alpha\beta\gamma}}}\ .
\eqn\volume
$$

According to the above interpretation, the theorem discussed in the previous
section guarantees that
$\f(x)$ represents the geodesic field of the membrane.
Note that eq.\volume\ implicitly suggests that we identify the term $g^2m$
with
the surface tension $\r$ of the membrane. That this is indeed the case can
be verified directly by inserting the solution \fslope\ into the action
\azionec . This operation yields the equivalent action for $X(\xi)$,
$$\eqalign{
S[X]&=-mg^2\int d^4x\sqrt{-{1\over 3!}\Phi_{\alpha\beta\gamma}
\Phi^{\alpha\beta\gamma}
\left(\,\int_Dd^3\xi\,\delta^{4)}\left(x-X(\xi)\right)\right)^2}\cr
&=-mg^2\int d^4x\int_Dd^3\xi\, \delta^{4)}\left(x-X(\xi)\right)
\sqrt{-{1\over 3!}{\dot X}_{\alpha\beta\gamma}{\dot
X}^{\alpha\beta\gamma}}\cr
&=-\r\int_Dd^3\xi
\sqrt{-{1\over 3!}{\dot X}_{\alpha\beta\gamma}{\dot
X}^{\alpha\beta\gamma}}
\ .\cr}
\eqn\sequiv
$$
which represents the action for a free membrane with an
{\it effective surface tension} $\r\equiv mg^2$.\foot{
Note that second term in \azionec\ does not contribute to \sequiv\ since
$\partial_\mu \J=0$}

Finally, as a consistency check, we wish to show that the gauge field
representation in terms of $F_{\mu\nu\rho}$ leads to the classical
equations of motion $\moto$.
To this end, note that $F_{\mu\nu\rho}$ satisfies the Bianchi identities
everywhere, so that in view of \volume\
$$
\partial_{\,[\lambda}\Pi_{\mu\nu\rho]}(x)=0
\eqn\bianchi
$$
at each spacetime point.
Then we can project eq.\bianchi\ along the membrane history, that is, we
evaluate $\Pi_{\mu\nu\rho}(x)$ at $x=X(\xi)$ and take the interior product
with ${\dot X}^{\lambda\mu\nu}$:
$$
\eqalign{
&{\dot X}^{\lambda\mu\nu}\partial_{\,[\lambda}\Pi_{\mu\nu\rho]}(\xi)=\cr
&\delta^{[abc]}\partial_a X^\lambda\partial_b X^\mu\partial_c X^\nu
\partial_{\,[\lambda}\Pi_{\mu\nu\rho]}(\xi)=\cr
&\delta^{[abc]}\partial_b X^\mu\partial_c X^\nu
\partial_a\Pi_{\mu\nu\rho}(\xi)=0\ .\cr}
\eqn\moto
$$
The last line in \moto\ is just the classical equation of motion
$\moto$ of the membrane.

\underbar{\it open membrane}

In the following we first derive a representation of an open membrane in
terms of a {\it gauge field} and then show that such a representation is
equivalent to the {\it geometric} representation of an open membrane in
terms of the Nambu-Goto action. The key ingredient of this equivalence is
the de Rham relation (4.8) with $p=2$ and our immediate objective is to
show how to derive the de Rham relation from an appropriate extension of
the action (4.10). More specifically, we need to modify the action (4.10) so
that the variation of the $B$-field yields an equation of the type:
$\dis{ \partial W={\rm const.}\times\hbox{closed string current}}$. With
this result in hands, property d), discussed in the previous subsection,
guarantees that we can construct a solution of the
type: $\dis{ W={\rm const.}\times \hbox{open \m\ current}}$. Using this
solution, we can finally transform the ``~gauge field~'' action into the
Nambu-
Goto geometric action associated with the open membrane.

The new action implementing the above idea is
$$\eqalign{
\so=&-g^2\int d^4x\,\sqrt{-{1\over 3!}\w(x)\W(x)}+
{1\over 3!}\int d^4x\,\W(x)\partial_{\,[\mu}B_{\nu\rho]}(x)\cr
&+{f\over 2}\int d^4x\, B_{\nu\rho}(x)\cs^{\nu\rho}(x)
-\st\int_{\partial D}ds d\tau\sqrt{-{1\over2}{\dot X}^{\mu\nu}
{\dot X}_{\mu\nu}}\ ,\cr
\cs^{\mu\nu}(x)&\equiv
\int_{\partial D}ds d\tau\,\delta^{4)}\left(x-X(s,\tau)\right)
{\dot X}^{\mu\nu}\ .\cr}
\eqn\azio
$$
Note that the action depends now explicitly on $B$ because
of the coupling to the string current, whereas the action \azionec\
depends on $B$  only through its field strength .

Varying the action \azio\ with respect to $\b$, $\w$ and $X^\mu(s,\tau)$
we get the following set of field equations
$$
\partial_\mu\W(x)=f\cs^{\nu\rho}(x)\ ,
\eqn\pdue
$$
$$
g^2{\w(x)\over\sqrt{-{1\over
3!}W_{\alpha\beta\gamma}W^{\alpha\beta\gamma}}}
+F_{\mu\nu\rho}=0\ ,
\eqn\fdue
$$
$$
\delta^{[ik]}\partial_i\Pi_{\mu\nu}\partial_k X^\nu=
-{f\over2}F_{\mu\nu\rho}
{\dot X}^{\nu\rho}\ .
\eqn\stringa
$$
Equation \fdue\ is the same as \fmunu, and again relates $F$ to $W$;
according to \derham , eq.\pdue\ admits a special solution, say $\hat\W$,
which is proportional to the current of an {\it open membrane}
having the string as its only boundary [\Am]:
$$\eqalign{
&\hat\W=f J_{(2)}^{\mu\nu\rho} ,\quad J_{(2)}^{\mu\nu\rho}(x)=
\int_Dd^3\xi\,\delta^{4)}\left(x-Z(\xi)\right)\Z^{\mu\nu\rho}\ ,
\cr
&\partial_\mu J^{\mu\nu\rho}_{(2)}=\cs^{\nu\rho}\ .\cr}
\eqn\wopen
$$
Finally eq.\stringa\ describes the motion of the boundary under the
action of a generalized Lorentz force produced by $F$.

Now, the expression of $F$ in terms of the slope field becomes
$$
F_{\mu\nu\rho}=-g^2{\ki(x)\over\sqrt{-{1\over
3!}\Phi_{\alpha\beta\gamma}
\Phi^{\alpha\beta\gamma}}}
=-{1\over f}\Pi_{\mu\nu\rho}(x)\ ,
\eqn\fopen
$$
with an effective surface tension $\r\equiv g^2f$. From \fopen\
one recovers the property that
$$
F_{\mu\nu\rho}\left(x=X(\xi)\right)={1\over f}\Pi_{\mu\nu\rho}(\xi)\ ,
\eqn\fpi
$$
and therefore eq.\stringa\ can be written in the form:
$$
\delta^{[ik]}\partial_i\Pi_{\mu\nu}\partial_k X^\nu=
{1\over2}\Pi_{\mu\nu\rho}(\xi){\dot X}^{\nu\rho}\ .
\eqn\intlorenz
$$
{\it Equation \intlorenz\ is precisely the equation of motion of an open
membrane
coupled
to its boundary}. As a matter of fact, the same equation of motion can be
derived from the
equivalent action for $X(\xi)$ and $X(s,\tau)$
$$
S^{\rm eff.}=-\r\int_Dd^3\xi
\sqrt{-{1\over 3!}{\dot X}_{\alpha\beta\gamma}{\dot
X}^{\alpha\beta\gamma}}
-\st\int_{\partial D}ds d\tau
\sqrt{-{1\over2}{\dot X}^{\mu\nu}{\dot X}_{\mu\nu}}\ ,
\quad\r\equiv g^2f\ ,
\eqn\aperta
$$
which is obtained by inserting the solution  \pdue\ in the action \azio.
Note that in this case
the second and third term in \azio\ cancel against each other. The final
result \aperta\ describes an open \m\ ``~geometrically~'' coupled to its
own boundary. Note also that according to \aperta , the boundary represents
a
physical object possessing a dynamics of its own.
In this sense, eq.\aperta\ can be considered as the generalization of the
Chodos and Thorne action for the open string with massive end-points
[\chodos].

\def\R{\dot R}
\chapter{The Maxwell field of a spherical bubble}

As an application of the previous formalism, in this section we derive
the form of the {\it static} Maxwell field associated with a
spherical bubble. To this end,
the results of the previous sections can be summarized
by the following ``~recipe~'' to evaluate $F$: given a solution
of the Lagrange equations ( more precisely a one parameter family of
membrane world-histories ),
compute the corresponding volume momentum $\Pi(\xi)$;
then use the embedding equations $x=X(\xi)$ together with the equation of
motion to write $\Pi$ as a function of the spacetime coordinate $x$.

The embedding in Minkowski space of a closed, spherically symmetric
membrane can be parametrized as follows,
$$
\eqalign{
x^0&=X^0(\tau)\cr
x^1&=R(\tau)\sin\theta\cos\phi\cr
x^2&=R(\tau)\sin\theta\sin\phi\cr
x^3&=R(\tau)\cos\theta
\cr}
\eqn\polars
$$
where $0\le\theta\le\pi$, $0\le\phi< 2\pi$,
and $-\infty< \tau< \infty$.

The assumption of spherical
symmetry enables us to describe this physical system as a point-particle in
a two-dimensional curved mini-superspace [\As]. Indeed, the
angular variables can be integrated out
of the Nambu-Goto action to give
$$
S\equiv\int dx^0 L(R,\R)=-4\pi\r\int d\tau R^2\sqrt{({\dot X^0}){}^2-\R{}^2}\
,
\quad\dot{()}\equiv{d\over d\tau}\ .
\eqn \minis
$$
If $\displaystyle{Y^A\equiv( X^0,\,R\,)}$ are interpreted as local coordinates
in (1+1)-dimensions, then the action $(\minis)$ takes the familiar form
describing the motion of a point-particle moving in a curved spacetime
in which the fundamental tensor \superg\ reduces to the form
$$
{\cal G}_{AB}\equiv-
{1\over 2\r^2}{\partial L^2\over\partial\Y^A\partial\Y^B}=
(4\pi\r R^2)^2\eta_{AB}\ .
\eqn\minigi
$$
According to eq.\mom\ the ``~momentum~'' is now
$$
P_A\equiv-\r\, G_{AB}\Y^B\ ,
\eqn\minimom
$$
while the Legendre transform \hquadro\ leads to
$$
H^2=(2\r \Y^A)P_A-L^2=-G^{AB}P_AP_B\ .
\eqn\minipiquadro
$$
Furthermore, from the H-J equation \hjgen\ one recovers
the ``~mass-shell~'' relation
$$
-G^{AB}P_AP_B=\r^2\ .
\eqn\modulopi
$$
for a ``~pointlike particle~'', of mass $\r$, in superspace.

Note that since polar coordinates have non homogeneous dimensions,
the components of the tangent three-vector ${\dot X}^{\mu\nu\rho}$ are
no longer dimensionless and are explicitly given,
in the ``~proper time~'' gauge $X^0=\tau$, by,
$$
\eqalign{
{\dot X}^{012}&=R^2(\tau)\sin\theta\cos\theta\cr
{\dot X}^{031}&=R^2(\tau)\sin^2\theta\sin\phi\cr
{\dot X}^{023}&=R^2(\tau)\sin^2\theta\cos\phi\cr
{\dot X}^{123}&=R^2(\tau)\dot R\sin\theta\ .
\cr}
\eqn\velocita
$$
Then, eq.\minis\ gives

$$
L(R,\R)\equiv -\r\sin\theta R^2\sqrt{1-\R^2}\
\eqn\minielle
$$
and the Lagrange equation leads to the following integral of motion
$$
\R^2=1-(R/R_0)^4  \eqn\erreditau
$$
where the constant $R_0$ represents the radius of the bubble
corresponding to the classical turning point $\R=0$.

{}From eq.\impa\ we can evaluate the components of the volume
conjugate momentum  for a generic bubble trajectory satisfying
eq.\erreditau\
$$
\eqalign{
\Pi_{021}&=\r{\cos\theta\over\sqrt{1-\R^2}}
=\r\left({R_0\over R}\right)^2\cos\theta\cr
\Pi_{013}&=\r{\sin\theta\sin\phi\over\sqrt{1-\R^2}}
=\r\left({R_0\over R}\right)^2\sin\theta\sin\phi\cr
\Pi_{032}&=\r{\sin\theta\cos\phi\over\sqrt{1-\R^2}}
=\r\left({R_0\over R}\right)^2\sin\theta\cos\phi\cr
\Pi_{123}&=\r{\R\over\sqrt{1-\R^2}}
=\r\left({R_0\over R}\right)^2\R=\r\left({R_0\over R}\right)^2
\sqrt{1-(R/R_0)^4}\ .
\cr}
\eqn\impulsi
$$
The above components satisfy the momentum
constraint $-(1/3!)\Pi_{\mu\nu\rho}\Pi^{\mu\nu\rho}=\r^2$.

Finally, in order
to obtain the corresponding Maxwell field, we recall that

$\displaystyle{r\equiv\sqrt{(x^1){}^2+(x^2){}^2+(x^3){}^2}\rightarrow R}$ on
the
membrane; thus, in order to have a static solution we have to replace
$R(\tau)$ with $r$ and express $\R$ in terms of $r$ consistently with
\erreditau.
Then, according to eq.\volume:
$$
\eqalign{
\Pi_{012}&\rightarrow
F_{012}={\r R_0\over m}{x^3\over r^3}\cr
\Pi_{013}&\rightarrow
F_{013}=-{\r R_0\over m}{x^2\over r^3}\cr
\Pi_{023}&\rightarrow
F_{023}={\r R_0\over m}{x^1\over r^3}\cr
\Pi_{123}&\rightarrow
F_{123}=
-{\r\over m}\left({R_0\over r}\right)^2\sqrt{1-\left({r/R_0}\right)^4}\ .
\cr}
\eqn\forze
$$
The Kalb-Ramond field $F_{\mu\nu\rho}$ defined above
satisfies the H-J equation \hjinx
$$
-{1\over 3!}F_{\mu\nu\rho}F^{\mu\nu\rho}={\r^2\over m^2}\ .
\eqn\effequadro
$$

To show that
$F\equiv(1/3!)F_{\mu\nu\rho}\,dx^{\mu}\wedge dx^{\nu}\wedge dx^{\rho}$
is a closed form, let us first set all the unessential constants equal
to one; then
$$\eqalign{
dF_{012}\, dx^0\wedge dx^1\wedge dx^2
&=-{1\over r^3}\left(1-{3z^2\over r^2}\right)dx^3\wedge dx^0\wedge dx^1
\wedge dx^2\ ,
\cr
dF_{013}\,dx^0\wedge dx^1\wedge dx^3
&={1\over r^3}\left(1-{3y^2\over r^2}\right)dx^2\wedge dx^0\wedge dx^1
\wedge dx^3\ ,
\cr
dF_{023}\,dx^0\wedge dx^1\wedge dx^3
&=-{1\over r^3}\left(1-{3x^2\over r^2}\right)dx^1\wedge dx^0\wedge dx^2
\wedge dx^3\ ,
\cr
dF_{123}\,dx^1\wedge dx^2\wedge dx^3 &=0
\cr}
\eqn\dieffe
$$
where,
in the last formula, we have taken advantage of the time independence of
$r$.
Collecting the above results we obtain
$$
dF=r^{-3}\left({3r^2-3\left[(x^1){}^2+(x^2){}^2+(x^3){}^2\right]\over r^2}
\right)
dx^0\wedge dx^1\wedge dx^2\wedge dx^3\equiv 0\ .
\eqn\effechiuso
$$
Finally, we notice that the field $\f$ has not been extended to the
whole spacetime, but only inside the region $0\le r\le R_0$, as it
is manifest from the expression $F_{123}$ in eq.\impulsi.
This field representation of a membrane is the counterpart of the Kastrup-
Rinke
spinning-string field which is defined only inside the cylindrical region
$r\le\pi A/2$ spanned by the classical string motion [\Au].

\chapter{Higher dimensional objects in higher dimensional spacetime}

To conclude our paper, in this section we present an extension of the
formalism discussed in Section 4 to the case of a hypersurface embedded in
a spacetime manifold with an arbitrary number of dimensions. This is partly
in recognition of the fact that the study of p-branes constitutes a relevant
part of the current research in the formal properties of strings and
membranes. For instance, membrane-like objects have been found recently
as solutions of
$d=10$, $N=2$ supergravity theories [\chs].
p-branes are objects
extended in {\it p} spatial dimensions and are defined in a
$D\ge p+1$ dimensional spacetime manifold ${\cal M}$
by assigning the pair $(U, X)$,
where $U$ is a connected, orientable, (p+1)-dimensional manifold
representing
the world hypersurface of the extended object, and $X$ is an embedding of
$U$ as a
timelike submanifold of ${\cal M}$. Then, the theory of
classical p-brane
dynamics is encoded into the generalized Dirac-Nambu-Goto action [\al]
$$
S=-\rho_p\int_U\sqrt{-{1\over
(p+1)!}\X^{\mu_1\dots\mu_D}\X_{\mu_1\dots\mu_D}}
\ ,\eqn\zero
$$
where $\rho_p$ is the hypersurface tension,
$\{\xi^1,\dots\xi^p\}$ are local coordinates on $U$, and
$$
\X^{\mu_1\dots\mu_D}=\partial_1 X^{\mu_1}\wedge\dots\wedge \partial_{(p+1)}
X^{\mu_D}\ ,
\eqn\uno
$$
represents the tangent (p+1)-vector to the world hypersurface.

Actually, one has to distinguish the case of p-branes with
$p+1<D$, from the extreme case of a ``~p-bag~'' characterized by $p+1=D$.
In the latter case the number of dimensions of the world-hypertube swept
by
the p-bag is equal to the number of spacetime dimensions of ${\cal M}$, so
that the embedding $X$ is equivalent to a {\it general coordinate
transformation} in ${\cal M}$. In other words, the gauge theory of a p-bag is
``~pure gauge~''. Nonetheless, its dynamics is not trivial.
\medskip
\underbar{$p+1<D$ {\it case}}

In this case the generalization of our formalism is almost straightforward
unless one requires some compactification
mechanism to get rid, at low energy, of the extra spatial dimensions. In this
connection, one usually considers the coupling of a p-brane to a
gauge (p+1)-form in the presence of gravity. While the actual process of
compactification is of no concern to us at present, it leads us to consider
the extension to higher dimensions of the
{\it general covariant} lagrangian model introduced in
Section 4.

For the sake of simplicity we shall consider here only closed
p-branes, and focus our discussion on two main points:

\noindent
i) the equivalence of the action
$$\eqalign{
S=&-\gb^2\int_{{\cal M}} d^Dx\sqrt{-g}
\sqrt{-{1\over (p+1)!}g_{\mu_1\nu_1}(x)\dots g_{\mu_{p+1}\nu_{p+1}}(x)
W^{\mu_1\dots\mu_{p+1}}(x)W^{\nu_1\dots\nu_{p+1}}(x)}\cr
&+{1\over (p+1)!}\int_{{\cal M}} d^Dx \sqrt{-g}\,
W^{\mu_1\dots\mu_{p+1}}\nabla_{[\,\mu_1}B_{\mu_2\dots\mu_{p+1}]}
\cr
&-{1\over 16\pi G_N}\int_{{\cal M}} d^Dx\sqrt{-g}\,R\cr
&F_{\mu_1\dots\mu_{p+1}}\equiv
\nabla_{[\,\mu_1}B_{\mu_2\dots\mu_{p+1}]}
\equiv \partial_{[\,\mu_1}B_{\mu_2\dots\mu_{p+1}]}\cr}
\eqn\seiuno
$$
with D-dimensional General Relativity minimally coupled with a p-brane;

\noindent
ii) the self-consistency of the model.

In eq.\seiuno, $\gb$ stands for the coupling constant while $g={\rm
det}g_{\mu\nu}(x)$; furthermore, $\nabla_\mu$ represents the usual,
Christoffel covariant derivative which, on account of the total
antisymmetry of the gauge field strength $F$, can be replaced by the
ordinary partial derivatives

The field equations derived from the action \seiuno\ are
$$
\partial_{\mu_1}\left[\sqrt{-g}\,W^{\mu_1\dots\mu_{p+1}}\right]=0\ ,
\eqn\seidue
$$
$$
-\gb^2{W_{\mu_1\dots\mu_{p+1}}\over\sqrt{-{1\over (p+1)!}
W^{\nu_1\dots\nu_{p+1}}W_{\nu_1\dots\nu_{p+1}}}}=
F_{\mu_1\dots\mu_{p+1}}\ ,
\eqn\seitre
$$
$$
R_{\mu\nu}-{1\over 2}g_{\mu\nu}R=8\pi G_N T_{\mu\nu}\ ,
\eqn\seiquattro
$$
where the energy momentum tensor is
$$\eqalign{
T_{\mu\nu}=&-2\left[{\gb^2\over 2p!}
{W_\mu{}^{\mu_2\dots\mu_{p+1}}W_{\nu\mu_2\dots\mu_{p+1}}\over
\sqrt{-{1\over
(p+1)!}W^{\mu_1\dots\mu_{p+1}}(x)W_{\mu_1\dots\mu_{p+1}}(x)}}
-B_{(\mu}{}^{\mu_3\dots\mu_{p+1}}\nabla_\tau
W_{\nu)}{}^\tau{}_{\mu_3\dots\mu_{p+1}}\right]\cr
&+g_{\mu\nu}L(W,B)\ .\cr}
\eqn\seicinque
$$
The indices $\mu$, $\nu$, in the second term in the square bracket are
symmetrized.
In complete analogy to the discussion of Section 4, the closed p-brane now
is represented by a solution of the covariant equation \seidue\
$$
W^{\mu_1\dots\mu_p}={m\over \sqrt{-g}}\int_U d^{p+1}\xi\,\delta^{(D)}
\left(x-
X(\xi)\right)\X^{\mu_1\dots\mu_{p+1}}=m\,J^{\mu_1\dots\mu_{p+1}}(x)
\ ,\eqn\seisei
$$
where the constant $m$ has now dimensions of $length^{(D-p-3)}$. To show
the
equivalence with D-dimensional General Relativity coupled to a closed
p-brane we need to identify equation \seicinque\ with the expression of the
energy-momentum tensor of the
extended object.  This is accomplished by noting that the
``~matter lagrangian~'' $L(W,B)$
vanishes for any solution of eq.\seitre, and so does the mixed $W$-$F$ term.
Therefore, the on-shell
energy-momentum tensor reads\foot{For conciseness, in the following we
define
$$
-{1\over (p+1)!}A_{\mu_1\dots\mu_{p+1}}A^{\mu_1\dots\mu_{p+1}}
\equiv ||A||^2\ .
$$
}%%%%%%%%%fine footnote
$$\eqalign{
T_{\mu\nu}&={\gb^2 m\over\sqrt{-g}}
{J_{\mu\mu_2\dots\mu_{p+1}}J_\nu{}^{\mu_2\dots\mu_{p+1}}\over
\sqrt{||J||^2}}\cr
&={\gb^2 m\over\sqrt{-g}}\int_U d^{p+1}\xi
{\X^\mu{}_{\mu_2\dots\mu_{p+1}}\X^{\nu\mu_2\dots\mu_{p+1}}
\over\sqrt{||\X||^2}}\delta^{(D)}\left(x-X(\xi)\right)\ ,\cr}
\eqn\seisette
$$
which proves our first point.

Next, as far as the self-consistency of the system of equations \seidue,
\seitre, \seiquattro, is concerned, we must check that the on-shell energy
momentum tensor is covariantly conserved.
$$\eqalign{
&\nabla_\mu T^{\mu\nu}=\cr
&{\gb^2 m\over\sqrt{-g}}\int_U d^{p+1}\xi \,
{\X^{\mu\mu_2\dots\mu_{p+1}}\X^\nu{}_{\mu_2\dots\mu_{p+1}}
\over\sqrt{||\X||^2}}\nabla_\mu\delta^{(D)}\left(x-X(\xi)\right)=\cr
&-{\gb^2 m\over\sqrt{-g}}\int_U d^{p+1}\xi\,\delta^{[ma_2\dots a_{p+1}]}
{\partial_{a_2}X^{\mu_2}\dots \partial_{a_{p+1}}X^{\mu_{p+1}}
\X^\nu{}_{\mu_2\dots\mu_{p+1}}
\over\sqrt{||\X||^2}}\nabla_m\delta^{(D)}\left(x-X(\xi)\right)=\cr
&-{\gb^2 m\over\sqrt{-g}}\int_U d^{p+1}\xi\,\delta^{[ma_2\dots a_{p+1}]}
\Biggl[\nabla_m\left(
{\partial_{a_2}X^{\mu_2}\dots \partial_{a_{p+1}}X^{\mu_{p+1}}
\X^\nu{}_{\mu_2\dots\mu_{p+1}}
\over\sqrt{||\X||^2}}\delta^{(D)}\left(x-X(\xi)\right)\right)\cr
&-\nabla_m\left({\partial_{a_2}X^{\mu_2}\dots
\partial_{a_{p+1}}X^{\mu_{p+1}}
\X^\nu{}_{\mu_2\dots\mu_{p+1}}
\over\sqrt{||\X||^2}}\right)\delta^{(D)}\left(x-X(\xi)\right)\Biggr]\ .\cr}
\eqn\seiotto
$$
The surface term in \seiotto\ does not contribute because the p-brane has
no boundary, and totally anti-symmetrized covariant derivatives can be
replaced by ordinary partial derivatives:

$$\eqalign{
\nabla_\mu T^{\mu\nu}&=
{\gb^2 m\over\sqrt{-g}}\int_U d^{p+1}\xi\,\delta^{[ma_2\dots a_{p+1}]}
\partial_m\left({\partial_{a_2}X^{\mu_2}\dots
\partial_{a_{p+1}}X^{\mu_{p+1}}
\X^\nu{}_{\mu_2\dots\mu_{p+1}}
\over\sqrt{||\X||^2}}\right)\delta^{(D)}\left(x-X(\xi)\right)\cr
&={1\over\sqrt{-g}}\int_U d^{p+1}\xi\,\left[
\delta^{[ma_2\dots a_{p+1}]}\partial_m\Pi^\nu{}_{\mu_2\dots\mu_{p+1}}
\partial_{a_2}X^{\mu_2}\dots \partial_{a_{p+1}}X^{\mu_{p+1}}\right]
\delta^{(D)}\left(x-X(\xi)\right)\cr
&=0\ .\cr}
\eqn\seinove
$$
Thus, the energy momentum tensor is conserved, and can be substituted into
the Einstein equations whenever $X(\xi)$ represents a solution of the
p-brane classical equation of motion
$$
\delta^{[ma_2\dots a_{p+1}]}\partial_m\Pi^\nu{}_{\mu_2\dots\mu_{p+1}}
\partial_{a_2}X^{\mu_2}\dots \partial_{a_{p+1}}X^{\mu_{p+1}}=0\ .
\eqn\seidieci
$$
In this connection, it is worth observing that, just as in the usual
formulation of General Relativity in four dimensions, the
classical equations of motion for the extended object emerge as a
consistency
condition on the general covariant formulation of the model. This is in
contrast to the formulation in Minkowski spacetime discussed in Section 4
where the equations of motion of the membrane were obtained by projecting
the Bianchi Identity
for the membrane field strength along the world-hypertube.

\medskip
\underbar{$p+1=D$ {\it case}}

The theory of a $D-1$-brane in a $D$-dimensional spacetime is
``~trivial~'' in the same sense as electrodynamics is trivial
in two dimensions, or three-form electrodynamics in four dimensions:
the field strength of the gauge potential, or , in our case the
Maxwell tensor associated to the extended objects, has as many indices
as the number of spacetime dimensions and thus must be proportional to the
Levi-Civita tensor in D-dimensions. However, even though such generalized
Maxwell tensors do not provide propagating degrees of freedom, their
gravitational interaction is non trivial. For instance, the three index
potential $A_{\mu\nu\rho}(x)$ in four dimensions gives rise to a background
energy density which enters the Einstein equations as an {\it effective
cosmological constant,} promoting this parameter to the role of dynamical
variable, and making it susceptible to a dynamical quantum adjustment to
zero [\haw]. Furthermore, there is a static effect that arises
whenever the gauge potential is coupled to a suitable current. Indeed, by
definition, a $D-1$-brane is an {\it open}
object, and according to our scheme we have to couple the associated
gauge potential to the current of the boundary, as in the following action
$$
\eqalign{
S=&-\gb^2\int_{{\cal M}}d^Dx\sqrt{-g}\sqrt{-{1\over
D!}W^{\mu_1\dots\mu_D}
W_{\mu_1\dots\mu_D}}+{1\over D!}\int_{{\cal M}}d^Dx\sqrt{-g}
W^{\mu_1\dots\mu_D}\nabla_{[\mu_1}B_{\mu_2\dots\mu_D]}\cr
&+{f\over (D-1)!}\int_{{\cal M}}d^Dx\sqrt{-g}
B_{\mu_2\dots\mu_D}J^{\mu_2\dots\mu_D}
-{1\over 2\pi\alpha'_D}\int_{\partial U} d^{(D-1)}\xi\sqrt{-{1\over (D-1)!}
\X^{\mu_2\dots\mu_D}\X_{\mu_2\dots\mu_D}}\cr
&-{1\over 16\pi G_N}\int_{{\cal M}}d^Dx\sqrt{-g}R\ .\cr}
\eqn\seiunouno
$$
The corresponding field equations are
$$
\partial_{\mu_1}\left[\sqrt{-g}W^{\mu_1\dots\mu_D}\right]=0\ ,
\eqn\seiunodue
$$
$$
\gb^2 {W_{\mu_1\dots\mu_D}\over\sqrt{||W||^2}}+F_{\mu_1\dots\mu_D}=0\
,
\eqn\seiunotre
$$
$$
\delta^{[a_1\dots a_{D-1}]}\partial_{a_1}\Pi_{\mu\mu_3\dots\mu_D}
\partial_{a_2}X^{\mu_3}\dots\partial_{a_2}X^{\mu_D}=-{f\over (D-1)!}
F_{\mu\mu_2\dots\mu_D}\X^{\mu_2\dots\mu_D}\ .
\eqn\seiunoquattro
$$
Equation \seiunodue\ can be solved as follows: in
$D$-dimensions a totally anti-symmetric tensor is necessarily proportional
to the Levi-Civita tensor, i.e.
$W^{\mu_1\dots\mu_D}=\epsilon^{\mu_1\dots\mu_D}
w(x)$. Accordingly, if we define the {\it dual boundary current} as
$$
J^*_{\mu_1}=\epsilon_{\mu_1\mu_2\dots\mu_D}J^{\mu_2\dots\mu_D}
=n_{\mu_1}\delta_{\partial U}^{(D-1)}(x)\ ,
\eqn\seiunocinque
$$
where $n_{\mu_1}$ represents the outward unit normal to the bag boundary
$\partial U$, and the boundary-delta function is defined by the relationship
$$
\delta_{\partial U}^{(D-1)}(x)=
\int_{\partial U} d^{(D-1)}\xi\,\delta^{(D-1)}\left[x-X(\xi)\right]\ .
\eqn\seiunosei
$$
Then, $\delta_{\partial U}^{(D-1)}(x)$  satisfies the following properties:
$$
\eqalign{
&\int_{{\cal M}}d^Dx\,\delta_{\partial U}^{(D-1)}(x)F(x)=
\int_{\partial U}d^{(D-1)}x\,\delta_{\partial U}^{(D-1)}(x)F(x)\ ,\cr
&\partial_\mu\Theta_U(x)=-n_\mu\delta_{\partial U}^{(D-1)}(x)\ ,\cr
}
\eqn\seiunosette
$$
where, $\Theta_U(x)$ is the {\it characteristic function} (or generalized
step function) of the open subset
$U$ of the spacetime manifold ${\cal M}$ associated with the bag interior
region.

Thus, eq.\seiunodue\ becomes
$$
\partial_\mu w(x)=-{f\over (D-1)!}J^*_{\mu_1}\Longrightarrow
w(x)=-f\Theta_U(x)+c\ ,
\eqn\seiunosette
$$
where $c$ is an arbitrary integration constant. Similarly, the generalized
Maxwell tensor can be expressed as follows
$$
F_{\mu_1\dots\mu_D}=-\gb^2 \epsilon_{\mu_1\mu_2\dots\mu_D}\
\eqn\seiunootto
$$
and we conclude that the bag field strength represents simply a constant
energy background
localized inside the bag .The coupling to gravity turns such a background
field into an {\it effective cosmological constant}. In fact, inserting the
two solutions \seiunosette,\seiunootto\ into the action \seiunouno\ we
obtain
$$
\eqalign{
S&=
-{1\over 2\pi\alpha'_D}\int_{\partial U} d^{(D-1)}\xi\sqrt{-{1\over (D-1)!}
\X^{\mu_2\dots\mu_D}\X_{\mu_2\dots\mu_D}}
-{1\over 16\pi G_N}\int_{{\cal M}}d^Dx\sqrt{-g}\left[R-
2\Lambda\Theta_U(x)
\right]\cr
&=-{1\over 2\pi\alpha'_D}\int_{\partial U} d^{(D-1)}\xi\sqrt{-{1\over (D-
1)!}
\X^{\mu_2\dots\mu_D}\X_{\mu_2\dots\mu_D}}-f\gb^2
\int_U d^D\xi\sqrt{-{1\over D!}
\X^{\mu_1\dots\mu_D}\X_{\mu_1\dots\mu_D}}\cr
&-{1\over 16\pi G_N}\int_{{\cal M}}d^Dx\sqrt{-g}R\ ,\cr}
\eqn\seiunonove
$$
which describes General Relativity, with an effective cosmological term
$\displaystyle{\Lambda=-8\pi G_N f\gb^2}$, inside the open spacetime
subset $U$.
Finally, the ``Lorentz force equation'' \seiunoquattro\ becomes the
boundary equation of motion subject to the ``~external force~'' provided by
the bag.

\medskip
Having established the self consistency of the matter-gravity sector of
the model in both the above cases, one can add gauge $(p+1)$ forms and
start to look
for compactified classical solutions describing the four dimensional
physical spacetime times
a compact, un-observable extra space. In this connection, the main technical
achievement
of our reformulation of p-brane theory, is that instead of working with
an hardly tractable system of equations coupling extended objects to local
fields, we can now deal with {\it local field equations} alone, thereby
greatly
simplifying the problem. Finally, it should be noted that the special case of
string induced compactification
in the framework of the Nambu-Hosotani field theory of strings, has been
discussed in ref. [\freund]. Our present generalization of the Nambu-
Hosotani
description of string dynamics provides the tool to investigate the problem
of higher dimensional object induced spacetime compactification.

\chapter{Aknowledgements}

One of the authors (E.S.) wishes to thank Prof.F.Legovini for many
discussions
about the subject of the paper.

\chapter{Appendix: the rank of a differential form}
%\centerline

The following definitions and theorems has been collected from ref.[\godb].

Let $\alpha$ be a p-form. A vector $X$ defines a (p-1)-form through
interior product $i(X)\alpha$.

Example.-
$\alpha={1\over 2}\alpha_{ij}dx^i\wedge dx^j$, and $X=X^k\partial_k$ then
$i(X)\alpha=X^i\alpha_{ij}dx^j$.

Given a subspace $F$ of the  manifold $E$, let us consider the set
$\Lambda(E/F)$ of all forms orthogonal to $F$
$$
\Lambda(E/F)=\{\alpha\in\Lambda(E):i(X)\alpha=0, \forall X\in F\}
$$
where $\Lambda(E)$ is the set of all forms on $E$.
\medskip
{\bf Def.} Let $\alpha$ be an exterior form on $E$.
The {\it associated vector subspace} $A(\alpha)$ is the greatest
subspace $H$ of $E$ such that $\alpha\in\Lambda(E/H)$, i.e. it is the
vector subspace
$$
\{X\in E:i(X)\alpha=0, \}\ .
$$
\medskip
{\bf Example.- } For a 2-form
$\alpha={1\over 2}\alpha_{ij}dx^i\wedge dx^j$ defined
over an n-dimensional manifold $E$, $i(X)\alpha=0\Rightarrow
v^i\alpha_{ij}=0,\ \forall j$. These equations define a subspace $H$ of the
original space $E$, spanned by the null eigenvectors of the $n\times n$
antisymmetric matrix $\alpha_{ij}$. The {\it rank} of this matrix is given
by the codimension of the associated space:
$$
\hbox{rank of}\ \alpha=n-\hbox{dim}H\ . \eqn\auno
$$
\smallskip
Generalizing this Example.-  we have:
\smallskip
{\bf Def.} The {\it associated system } of $\alpha$ is the subspace
$\Lambda^*(\alpha)=A(\alpha)^{\perp}$ of the forms orthogonal to the
space $A(\alpha)$.
\smallskip
{\bf Def.} The {\it rank} of $\alpha$ is the dimension of the associated
system $\Lambda^*(\alpha)$.
\smallskip
So, eq.\auno\ is true in general, and we can say that
$$
\eqalign{\hbox{rank}&=\hbox{least number of linearly independent forms
necessary to express}\ \alpha\cr
&=\hbox{codimension of the associated space}\ A(\alpha)\ .
\cr}
$$
Also, the following inequality holds true:
 $\hbox{form-degree}\le \hbox{rank}\le\hbox{manifold-dimension}$; thus
$$
\eqalign{
&\hbox{rank ( 0-form )}=0\cr
&\hbox{rank ( null-form )}=0\cr
&\hbox{rank (non-vanishing 1-form )}=1
\cr
&\hbox{degree}\> n\ \Rightarrow\  \hbox{rank}\> n\cr
&\hbox{degree}\> n-1\ \Rightarrow\ \hbox{rank}\> n-1\cr
&\hbox{degree}\> n-2\ \Rightarrow\  \hbox{either}\> n-2 \hbox{or} n\cr
&\hbox{degree}\> 2\ \Rightarrow\ \hbox{rank even}=2s\ \hbox{if}\
\alpha^s\ne 0\ \hbox{and}\ \alpha^{s+1}=0
\cr}
$$
{}From the above relations one sees that a {\it 3-form in four dimensions has
always rank three}, which is the property we have used throughout the
paper.

\smallskip
{\bf def.} The {\it characteristic subspace} of $\alpha$ at a point $y$ of
$E$, is the subspace $C_y(\alpha)$ of the tangent space $T_y(E)$,
that is,   the intersection
of $\Lambda(\alpha(y))$ and $\Lambda(d\alpha(y))$
\smallskip
{\bf def.} The {\it characteristic system } of $\alpha$ at a point $y$ of
$E$, is the subspace $C^*_y(\alpha)$ of the cotangent space $T^*_y(E)$,
orthogonal to $C_y(\alpha)$
\smallskip
{\bf def.} The {\it class} of $\alpha$ is the  dimension of the
characteristic system.
\smallskip
For a closed form, class=rank.
\smallskip
{\bf Example.- } $\alpha=(x^2+y^2)dy$ on the real plane $R^2$;
$d\alpha=2x\, dx\wedge dy$. Then, rank of $\alpha= 1$, unless $x=y=0$.
If $x\ne 0$, then
class$=2$; if $x=0$, then $y\ne 0$ class$=1$; if $x=y=0$ then
class$=0$.
\smallskip
The constant(=is the same over the whole
manifold) class of the form $\alpha$ is the least  number of independent
functions necessary to express $\alpha$.
\smallskip
{\bf Example.- } $\beta=\beta(x,y,z)dx\wedge dy$ in $R^3$
is of rank 2 and class 3
(in three dimensions).
\smallskip
If degree=constant class,  then there exists a system of local coordinates
$(y_1,\dots,y_m)$, on an open set $U$, such that:
$$
\alpha=dy_1\wedge\dots\wedge dy_m\qquad\hbox{on}\quad U
$$
\medskip
\centerline{Summary}

A non-vanishing $(n-1)$-form defined over an $n$-dimensional space  has
rank $n-1$ ( if it is non-vanishing ).

If the form is not closed, its class is $n$.

If it is closed, its class is $n-1$

If constant class=degree, then rank=degree and
$$\alpha=\alpha(y_1,\dots y_p)dy_1\wedge
\dots\wedge dy_p=dz_1\wedge\dots\wedge dz_p\ ,
$$
where
$z_1=\int dy_1\alpha(y_1,\dots y_p)$ and $z_j=y_j$ for $y>1$.
\medskip
\chapter{Appendix B: the canonical H-J action \hamact\ }

Equation \hamact\ can be derived from the Hamilton principle in
parametrized
form.  Let us start from the canonical action
$$
S[X,\Pi]=\int_{X_1(s)}^{X_2(s)} d^3\xi \left[{1\over 3!}\P
{\dot X}^{\lambda\mu\nu}-h(X,\Pi)\right]\ ,\quad h(X,\Pi)={1\over 3!}\P\p\ .
\eqno (B1)
$$
The action (B1) is not invariant under re-parametrization and must be
parametrized ``~by hand~''[\kuchar],
which means introducing three new parameters
$\dis{\{\sigma^a\}=\{\sigma^0,\sigma^1,\sigma^2\}}$, describing the path
of the \m\ in state space(~= phase space $\times$ parameter space~).
Moreover, we promote the original parameters
$\xi^a$ to the role of dynamical variables, i.e. the new configuration
variables are $\dis{\{X^\mu(\sigma),\xi^a(\sigma)\}}$. According
to this new parametrization, the action (B1) reads
$$
S[X,\Pi;\xi]=
\int_D d^3\sigma\left[{1\over 3!}\P{\dot X}^{\lambda\mu\nu}-
\sqrt{-\gamma}h(X,\xi,\Pi)\right]\ ,
\eqno (B2)
$$
where
$$
\sqrt{-\gamma}\equiv {\rm
det}\Big\vert{\partial\xi^i\over\partial
\sigma^a}\Big\vert=\delta^{[abc]}\delta_{[ijk]}\partial_a\xi^i
\partial_b\xi^j\partial_c\xi^k\ .
\eqno (B3)
$$
The actions (B1) and (B2) are numerically equivalent, so that variation
with respect to the canonical variables $X^\mu$ and $\P$ yields the same
equations of motion. The additional variation with respect to $\xi^a$
gives the rate of change of the Hamiltonian along the \m\ world-track
$$
\delta^{[abc]}\delta_{[ijk]}\partial_b\xi^j\partial_c\xi^k
{\partial h\over\partial\sigma^a}=\sqrt{-\gamma}
{\partial h\over\partial\xi^i}\ .
\eqno (B4)
$$
However, the new parametrization of the action (B1) is by no means trivial
since the canonical form of the action with respect to the variables
$\xi^a$ leads us to a new vanishing hamiltonian, which is the distinguishing
feature of a reparametrization invariant theory. In fact,
having enlarged the configuration space, it follows that we can also
define  the corresponding tangent element $\dis{\{ {\dot X}^{\mu\nu\rho},
{\dot \chi}^{ijk}\}}$ where
$$
{\dot \chi}^{ijk}\equiv \delta^{[abc]}\partial_a\xi^i
\partial_b\xi^j\partial_c\xi^k\ .
\eqno (B5)
$$
By setting
$$
\Pi_{ijk}\equiv -\delta_{[ijk]}h
\eqno (B6)
$$
we can write eq. (B2) in the suggestive form
$\dis{S=\int_D d^3\sigma{1\over 3!}{\bf\Pi}\cdot{\bf\dot X}}$,
where ${\bf \Pi}\equiv \{\P\ ,\Pi_{ijk}\}$ and
${\bf\dot X}=\{{\dot X}^{\mu\nu\rho}\ , {\dot \chi}^{ijk}\}$. However, the
momenta ${\bf \Pi}$ cannot be varied freely because the components
$\Pi_{ijk}$ are related
by the duality (~in parameter space~) relation (B6) to $h(X\ ,\xi\ , P)$.
Therefore, we
must incorporate the constraint (B6) into the action by means of a lagrange
multiplier
$N^{ijk}$ :
$$
S[X,\Pi,N;\xi]={1\over 3!}
\int_Dd^3\sigma\left[\P{\dot X}^{\lambda\mu\nu}+
\Pi_{ijk}{\dot \chi}^{ijk}+N^{ijk}\left(\Pi_{ijk}+
\delta_{[ijk]}h(X,\xi,\Pi)\right)\right]\ .
\eqno (B7)
$$
Varying $\xi$ in (B7) we obtain now the equation of motion
$$
\delta^{[abc]}\partial_a \xi^l\left(\partial_l \Pi_{ijk}\right)
\partial_b \xi^j\partial_c \xi^k=0\ .
\eqno (B8)
$$
But, $\Pi_{ijk}$ is  a rank-three, totally anti-symmetric tensor in
parameter
space, and so it can be written as $\dis{\Pi_{ijk}\equiv
\delta_{[ijk]}\Pi(\xi)}$.
Then, eq. (B8) gives
$$
\partial_l \Pi(\xi)=0\,\Rightarrow \Pi(\xi)={\rm const.}\equiv\r^2\ .
\eqno (B9).
$$
Varying $\Pi_{ijk}$ in (B7) we obtain the classical solution for the
Lagrange multiplier
$$
N^{ijk}_{\rm cl.}={\dot \chi}^{ijk}\ ,
\eqno (B10)
$$
which shows that $d^3\sigma\,\delta_{[ijk]}N^{ijk}_{\rm cl.}$ is an {\it
invariant measure} of the \m\ proper volume.

Note that $h(P)$ does not depend from $\xi^a$, so we are allowed to use
the equations of motion to eliminate $\xi^a$ from the action.

By inserting solutions (B9),(B10) in eq. (B7),  and defining $\dis{ N\equiv
{1\over 3!}\delta_{[ijk]}N^{ijk}}$, the Hamilton-Jacobi action \hamact\
is obtained by discarding the boundary term $\int_Dd^3\sigma N(\sigma)$
representing the proper volume of the world-track between the initial
and final \m.

%%%%%%%%%
\refout
\bye